\documentclass[journal,letterpaper,twoside,twocolumn]{IEEEtran}
\usepackage{amsmath,amssymb,graphicx,psfrag,cite}
\usepackage[normalem]{ulem} 
\usepackage[usenames,dvipsnames]{color} 
\usepackage{balance}
\usepackage{mathrsfs}   
\usepackage{paralist}

\graphicspath{{figure/}}

\usepackage{color}

\definecolor{emphasis}{RGB}{204,0,204}
\renewcommand{\uline}[1]{{\color{emphasis}#1}}

  \renewcommand{\sout}[1]{} \renewcommand{\uline}[1]{#1} 

\title{Hierarchical Distribution Matching for Probabilistically Shaped Coded Modulation} 
\author{Tsuyoshi~Yoshida,~\IEEEmembership{Member,~IEEE,} Magnus~Karlsson,~\IEEEmembership{Fellow,~OSA;~Senior~Member,~IEEE,} \\ and Erik~Agrell,~\IEEEmembership{Fellow,~IEEE} 
}%
\begin{document}
\IEEEspecialpapernotice{(Invited Paper)}
\maketitle

\begin{abstract}

The implementation difficulties of combining distribution matching (DM) and dematching (invDM) for probabilistic shaping (PS) with soft-decision forward error correction (FEC) coding can be relaxed by reverse concatenation, for which the FEC coding and decoding lies inside the shaping algorithms. PS can seemingly achieve performance close to the Shannon limit, although there are practical implementation challenges that need to be carefully addressed. We
propose a hierarchical DM (HiDM) scheme, having fully parallelized input/output interfaces and a pipelined architecture that can efficiently perform the DM/invDM without the complex operations of previously proposed methods such as constant composition DM (CCDM). Furthermore, HiDM can operate at a significantly larger post-FEC bit error rate (BER) for the same post-invDM BER performance, which facilitates simulations. These benefits come at the cost of a slightly larger rate loss and required signal-to-noise ratio at a given post-FEC BER.
\end{abstract}


\section{Introduction}
\label{sec:intro}

Multilevel modulation formats have been intensively investigated in coherent optical communications due to the growing traffic demands and requirements for high spectral efficiency. To relax the signal-to-noise ratio (SNR) requirements for such formats, two independent approaches can be employed. The first is {\em coding}, or forward error correction (FEC), where hard- and soft-decision schemes for fiber-optic communications have demonstrated bit error rates (BER) as low as $10^{-15}$ \cite{chang_2010_cm, agrell_2018_ipc}. The second is {\em shaping}, which aims at reducing the average signal \sout{power} \uline{energy} by spherically confining the modulation levels in signal space. 

Two shaping approaches can be distinguished; geometric shaping, which uses uniform (equiprobable) modulation levels that are selected to be more or less spherically distributed, and probabilistic shaping (PS), which is based on using low-amplitude constellation points more frequently. Both schemes aim\sout{s} at approximating a Gaussian distribution, which is capacity-achieving for the Gaussian channel. The ultimate SNR improvement for a\sout{n} multidimensional quadrature amplitude modulation (QAM) format relative to the Gaussian channel capacity approaches $\pi e/6$ (1.53~dB) as the number of dimensions tends to infinity, as shown, e.g., by Forney and Wei \cite{forney_1989_jsac}, who also demonstrated a simple approach to create multidimensional shaped constellations. 
Calderbank and Ozarow pointed out that an equiprobable (uniform) multidimensional constellation could be viewed as a lower-dimensional nonequiprobable (nonuniform) signaling scheme \cite{calderbank_1990_tit}. 
Kschischang and Pasupathy \cite{kschischang_1993_tit} studied nonuniform signaling (probabilistic shaping), and showed that high shaping gains could be realized already for relatively low dimensions and with limited complexity based on Huffman codes. However, the varying bit rate and synchronization problems of such schemes may preclude their usefulness in practical systems. In \cite{raphaeli_2004_tcom}, these drawbacks were proposedly amended by keeping a constant bit-to-symbol rate at the expense of dropping (puncturing) bits. 


As coherent optical communication assisted by digital signal processing has been realized \cite{roberts_2009_jlt}, multidimensional signaling for optical links has regained interest, and efficient formats were proposed for the inherently four-dimensional optical signals \cite{agrell_2009_jlt}. For the nonlinear optic channel, variants of multidimensional geometric 
shaping \cite{millar_2014_oe,dar_2014_isit,shiner_2014_oe,kojima_2017_jlt} have been studied, as well as probabilistic shaping \cite{beygi_2014_jlt,yankov_2016_jlt}, based on the scheme proposed in \cite{raphaeli_2004_tcom}.

When combining coding and shaping, there are numerous issues to consider, e.g., the ordering of the schemes and the fact that the presence of one scheme may affect the performance of the other. While these issues were not discussed in the earlier works \cite{forney_1989_jsac,kschischang_1993_tit,agrell_2009_jlt,dar_2014_isit}, an outer FEC is often assumed \cite{shiner_2014_oe,kojima_2017_jlt}, or even necessary  \cite{raphaeli_2004_tcom,yankov_2016_jlt}. 

On the other hand, recently B\"ocherer \emph{et al.} \cite{bocherer_2015_tcom} introduced probabilistic amplitude shaping (PAS) based on a \emph{reverse concatenation} architecture, meaning that the FEC coding/decoding is performed inside the shaping algorithm, thus acting on nonuniformly distributed bits. This may realize improved performance and also enable rate adaptation in the shaping scheme rather the FEC, which may provide increased granularity. We note that reverse concatenation was indeed studied much earlier, in \cite{bliss_1981_ibm}, and the use of soft-decision FEC was also investigated \cite{fan_1999_globecom,djordjevic_2006_jlt}. Since its emergence, the PAS scheme in the transmitter and its termination in the receiver have been called distribution matching (DM) and distribution dematching (invDM), respectively, and we will use this terminology henceforth.

With the normal (nonreverse) concatenation architecture, invDM should be placed before soft-decision FEC decoding. Then soft demapping must convert the multidimensional fine-bit-resolution signals into logarithmic ratios of post-probabilities (\emph{a posteriori} log-likelihood ratios or L-values) \cite[Eq.~(3.31)]{bicmbook} for the decoder, which is prohibitively complex to implement for long codes. On the other hand, reverse concatenation changes the invDM input/output signals to binary sequences. This makes the implementation of invDM much simpler, and enables the number of dimensions (the code length) to be larger, leading to better performance.

As the DM for reverse concatenation PS, constant composition DM (CCDM) based on arithmetic coding was proposed in \cite{schulte_2016_tit} and examined for a fiber-optic channel in \cite{buchali_2016_jlt}. These state-of-the-art works show great performance by an almost ideal DM. However, there is room for simplification of these in high-throughput optical fiber communication systems.
DM and invDM consume nonnegligible circuit resources, and in practice there will \sout{always} be implementation penalties \uline{caused by imperfect signal processing used in deployable hardware at high throughputs,} relative to the achievable rate that assumes ideal DM and FEC. In addition, system performance monitoring must be reconsidered when we employ reverse concatenation PS; since the invDM may cause the BER to increase, post-invDM BER should be used instead of post-FEC BER.

In this paper, we propose and analyze in detail a DM scheme, which was briefly introduced in \cite{yoshida_2018_ecoc}. We compare it with CCDM. We also, for the first time, explain the shaped frame structure, show the result of the DM-to-invDM back-to-back error insertion test, and suggest a method to estimate the post-invDM BER from the post-FEC BER.


This paper is organized as follows: In Secs.~\ref{sec:system} and \ref{sec:monitor}, we describe the basic system model and relevant performance metrics. Implementation issues are discussed in Sec.~\ref{sec:impl}, and the proposed DM and framing are discussed in Sec.~\ref{sec:prop}. Simulation results comparing our proposed scheme with CCDM are described in Sec.~\ref{sec:sim}. The proposed error-rate estimation is presented in Sec.~\ref{sec:perfmon}. A summary of the work and a future outlook are provided in Sec.~\ref{sec:cncl}.

\section{System model}
\label{sec:system}
\begin{figure}
	\begin{center}
		\includegraphics[scale=0.42]{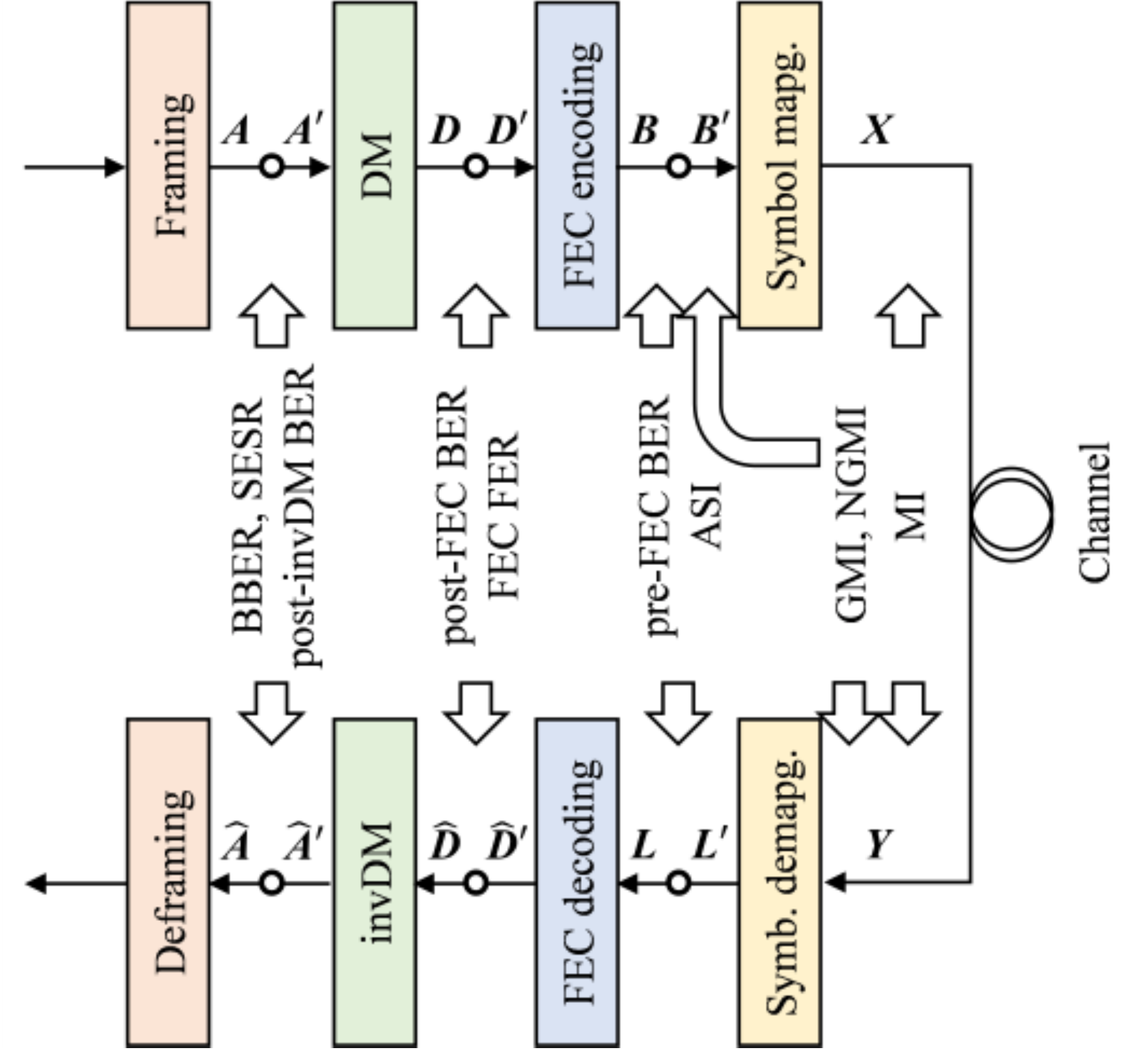} 
		\caption{System model including signal notation and performance metrics.}
		\label{fig:system}
	\end{center}
	\vspace{-0.6cm}
\end{figure}

Fig.~\ref{fig:system} shows the system model and the corresponding performance metrics to be discussed later. We \sout{assume the use of} \uline{consider shaping of one-dimensional} pulse amplitude modulation (PAM). However, its extension to two-dimensional QAM or optical four-dimensional modulation (two quadratures in two polarizations) is straightforward.

\uline{The notation is summarized in Tab.~\ref{tab:def_var_set}. Throughout the paper, we use  boldface symbols to denote vectors and calligraphic symbols to denote sets.} 
At the transmitter side, the incoming binary sequence is framed into blocks of $\boldsymbol{A} \in \mathcal{B}^{N_{\text{block}}}$ bits by following, e.g., the modern optical transport network (OTN) standard protocol \cite{ITU-T_G.709,ITU-T_G.8201} of $n \times 100 \text{G}$ optical transport units (OTUCn)\uline{.} \sout{, where $\mathcal{B} \in \{ 0, 1\}$ and $N_{\text{block}}$ is the block size of OTUCn, which is $130560 n$}. \sout{A bullet in Fig.~\ref{fig:system} indicates a length conversion.} In the DM, the binary sequence $\boldsymbol{A}' \in \mathcal{B}^{N_{\text{u}}}$ is converted to the shaped \sout{symbol} \uline{bit} sequence $\boldsymbol{D} \in$ \uline{$\mathcal{B}^{m N_{\text{s}}/2}$} including placeholders for FEC parity bits in the following block, where $m$ is the number of bits in \sout{the $2^m$-PAM}\uline{a QAM} symbol. \sout{$N_{\text{u}}$ is the number of (uniformly distributed and independent) DM input bits per DM word, 
and $N_{\text{s}}$ is the DM word length (or the DM output block length).} 
\uline{The bit sequence $\boldsymbol{D}$ }
\sout{corresponds to} \uline{determines} the \uline{absolute amplitudes of $N_{\text{s}}$} PAM symbol\uline{s.} \sout{sequence $\boldsymbol{X} \in \mathcal{X}^{N_{\text{s}}}$, where $\mathcal{X}$ denotes a PAM symbol.} A systematic binary FEC encoder generates $n_{\text{c}}-k$ (uniformly distributed) parity bits from $k$ payload bits $\boldsymbol{D}' \in \mathcal{B}^k$, and outputs the FEC codeword $\boldsymbol{B} \in \mathcal{B}^{n_{\text{c}}}$. The bits $\boldsymbol{B}' \in \mathcal{B}^{m}$ are mapped to the \sout{PAM} \uline{QAM} symbol \sout{$X$} \uline{$\boldsymbol{X}$} $\in$ \sout{$\mathcal{X}$} \uline{$\mathcal{X}^2$}\uline{, where $\mathcal{X}$ denotes a PAM symbol set}. \uline{A bullet in Fig.~\ref{fig:system} indicates a length conversion, e.g., $\boldsymbol{B}$ and $\boldsymbol{B}'$ have the same elements from $\mathcal{B}$ but have different lengths of $n_{\text{c}}$ and $m$ for interfacing, respectively,  the FEC encoding output and symbol mapping input.}

At the receiver side, the received \uline{QAM} symbol \sout{$Y$} \uline{$\boldsymbol{Y} \in \mathcal{R}^2$} is demapped by bit-metric decoding to \emph{a posteriori} L-values $\boldsymbol{L}' \in \mathcal{R}^{m}$ \cite[Eq.~(3.31)]{bicmbook}, where $\mathcal{R}$ denotes a \sout{continuous or a quantized} real number \uline{set}. Based on the L-values $\boldsymbol{L} \in \mathcal{R}^{n_{\text{c}}}$, FEC decoding recovers the payload bits $\widehat{\boldsymbol{D}}' \in \mathcal{B}^k$. The decoded bit sequence $\widehat{\boldsymbol{D}} \in$ \uline{$\mathcal{B}^{m N_{\text{s}} /2}$} is dematched to $\widehat{\boldsymbol{A}}' \in \mathcal{B}^{N_{\text{u}}}$, and finally $\widehat{\boldsymbol{A}} \in \mathcal{B}^{N_{\text{block}}}$ is deframed.


\begin{table}
	\caption{\uline{Definitions of \uline{notation}.}}
	\label{tab:def_var_set}
	\vspace{-0.4cm}
	\begin{center}
	\begin{tabular}{cl}
		\hline
		Parameter & Description \\
		\hline\hline
		$N_{\text{block}}$ & Block size of OTUCn, $130560 n$ \\
		$n$ & ``n'' in ``OTUCn'' \\
		$N_{\text{u}}$ & Number of DM input bits per DM word \\
		$N_{\text{s}}$ & DM word length (number of PAM symbols per DM word) \\
		$n_{\text{c}}$ & Number of information bits per FEC codeword \\
		$k$ & Number of parity bits per FEC codeword \\
		$m$ & Number of bits per QAM symbol \\
		$\mathcal{B}$ & $\{0,1\}$ \\
		$\mathcal{X}$ & $2^{m/2}$-ary PAM symbol set for even $m$ ($m \ge 2$) \\
		$\mathcal{R}$ & Real number set \\
		\hline
	\end{tabular}
	\end{center}
\end{table}

\section{Performance metrics and performance monitoring}
\label{sec:monitor}
\subsection{Achievable rates and metrics as post-FEC BER predictor}
Traditionally, the performance requirement for the optical layer has been a BER after FEC decoding down to $10^{-15}$ following standards for fiber communications \cite{ITU-T_G.975.1,agrell_2018_ipc}. More modern performance metrics are based on information theory, which quantifies achievable information rates (AIR\uline{s}) to \uline{lower-bound} the capacity. The mutual information of the transmitted symbol \sout{$X$} \uline{$\boldsymbol{X}$} and the received symbol \sout{$Y$} \uline{$\boldsymbol{Y}$}, \sout{$I(X;Y)$} \uline{$I(\boldsymbol{X};\boldsymbol{Y})$}, \uline{is such an AIR, which is achievable} \sout{can be a capacity lower bound} with nonbinary coding, bit-interleaved coded modulation (BICM) with iterative demapping, or multilevel coded modulation \cite{schmalen_2017_jlt}. For BICM without iterative demodulation, an AIR is given by the generalized mutual information (GMI) \cite[\sout{Sec.~4.3}\uline{Eq.~(4.55)}]{bicmbook}. The \sout{normalized} GMI \sout{(NGMI, GMI$/m$)} works as a good post-FEC BER predictor for uniform and independent pragmatic signaling \cite{alvarado_2015_jlt}. 
A practical, nonideal FEC has a rate loss, e.g., the required \sout{NGMI}\uline{FEC code rate $R_{\text{c}}$} for a post-FEC BER $<10^{-15}$ is \sout{larger}\uline{smaller} than the \sout{FEC code rate $R_{\text{c}}$}\uline{normalized GMI (NGMI) GMI$/m$} \cite[Tab. III]{alvarado_2015_jlt}.

For reverse concatenation PS with \uline{matched} bit-metric decoding \uline{(bmd)}, an AIR \uline{in bit per channel use (bpcu)} is \cite{bocherer_2015_tcom,bocherer_2017_arxiv}
\begin{IEEEeqnarray}{rCL}
	\label{eq:RBMD}
	R_{\text{bmd}}^{\text{ps}} &=& \uline{H(\boldsymbol{X})-\sum_{i=1}^{m} H(B_i \mid \boldsymbol{Y})} ,
\end{IEEEeqnarray}
where $H(\cdot)$ and $H(\cdot\,|\,\cdot)$ denote entropy and conditional entropy, resp.\uline{, and $B_i$ denotes a bit at bit level index $i$ before QAM symbol mapping.} $R_{\text{bmd}}^{\text{ps}}$ takes the same value as a \uline{modified expression of} \sout{the} GMI \sout{calculated for}\uline{applicable to} PAS \uline{defined in \cite[Eq.~(1)]{cho_2017_ecoc} with a matched auxiliary channel $q$, which was also shown in \cite[Sec.~III]{yoshida_2017_ptl}}. \sout{This} \uline{Eq.~\eqref{eq:RBMD} quantifies the rate as} \sout{can be} a common expression for pragmatic BICM and reverse concatenation PS.
While $R_{\text{bmd}}^{\text{ps}}$ is the rate with ideal FEC and ideal DM, the information rate \uline{in bpcu} with a practical FEC (discussed in the previous paragraph) is defined as
\begin{IEEEeqnarray}{rCL}
	\label{eq:info_rate}
	R & = & \uline{ \frac{N_{\text{u}}}{N_{\text{s}}/2} } \le \uline{H(\boldsymbol{X})}-\left( 1 - R_{\text{c}} \right) m  , 
\end{IEEEeqnarray}
where the symbol entropy \sout{$H(X)$} \uline{$H(\boldsymbol{X})$} is calculated from the probability mass function (PMF) of the transmitted \uline{QAM} symbol \sout{$X$}\uline{$\boldsymbol{X}$}, \sout{$P_{X}(x)$}\uline{$P_{\boldsymbol{X}}(\boldsymbol{x})$}, \sout{$x \in \mathcal{X}$}\uline{$\boldsymbol{x} \in \mathcal{X}^2$. Note that $H(\boldsymbol{X})$ is equivalent to $2H(X)$ in the case of PAS, where $X$ denotes a PAM symbol in $\mathcal{X}$}.
Non-ideal DM has a rate loss\uline{, i.e., difference between the entropy and the information rate, shown in \cite[Sec.~V-B]{bocherer_2015_tcom}, \cite[Eq.~(4)]{fehenberger_2018_tcom}, \cite[Eq.~(21)]{gultekin_2018_sitb} for PAS, which can be expressed as}
\begin{IEEEeqnarray}{rCL}
	\label{eq:rateloss}
	\uline{R_{\text{loss}}} &=& \uline{H(\boldsymbol{X})}-\left( 1 - R_{\text{c}} \right) m - \uline{\frac{N_{\text{u}}}{N_{\text{s}}/2}} \ge 0 
\end{IEEEeqnarray}
\uline{for reverse concatenation PS. The inequality is not restricted to PAS but more general. For certain (high performance) DMs\cite{schulte_2016_tit,fehenberger_2018_tcom,gultekin_2017_pimrc,gultekin_2018_isit,gultekin_2018_sitb,schulte_2018_arxiv}, the loss is mainly due to a finite (insufficient) length of the DM word.}

As post-FEC BER predictors for PS systems using QAM, the NGMI was proposed in \cite[Eq.~(6)]{cho_2017_ecoc}\sout{.} \uline{as} 
\sout{Using the conditional entropy of $B_i$ given $Y$ and generalizing to arbitrary constellations, it can be shown that the NGMI in} 
\sout{is a special case of}
\begin{IEEEeqnarray}{rCL}
	\label{eq:NGMI_PAS}
	\text{NGMI} &=& 1 - \uline{\frac{ 2 H(X) - \text{GMI} } {m} },
\end{IEEEeqnarray}
\uline{which means $\text{GMI}/m$ in a non-shaped case.}

The asymmetric information (ASI) \cite[Eq.~(11)]{yoshida_2017_ecoc_met,yoshida_2017_ptl}
\begin{IEEEeqnarray}{rCL}
	\label{eq:ASI}
	\text{ASI} &=& 1 - h (L_{\text{a}} \mid |L_{\text{a}}| ) \\
 		     &=& 1 - \frac{1}{m} \sum_{i=1}^{m} h(B_i \mid L_i)
\end{IEEEeqnarray}
was also proposed for the same purpose, where $h(\cdot)$ denotes differential (continuous) entropy. We denote the symmetrized \uline{\emph{a posteriori}} L-value, or asymmetric L-value, with $L_{\text{a}}=(-1)^{B}\cdot L$, where $B \in \mathcal{B}$ is a sample of $\boldsymbol{B}$, and $L \in$ \sout{$\mathcal{L}$} \uline{$\mathcal{R}$} is a sample of $\boldsymbol{L}$ corresponding to the sampled bit $B$. 
\sout{When the channel assumed in the soft-demapping is matched to the true channel, the ASI is equivalent to the NGMI} 

In practice, \sout{however, the} soft-demapping circuits operate with a finite bit resolution, which causes a minor performance loss. \sout{The GMI or NGMI cannot} \uline{Here GMI, NGMI, and ASI can} account for that loss because \uline{GMI is based on both} the received symbol \sout{$Y$}\uline{$\boldsymbol{Y}$} {and the auxiliary channel $q$} \cite[Sec.~III-C]{alvarado_2015_jlt} \sout{is used,  but} \uline{and} the ASI \sout{can because it} is based on the L-values just before the FEC decoder\sout{.}\uline{, while \eqref{eq:RBMD} does not account for it.} 
The post-FEC BER can be estimated from the NGMI or the ASI. However, since the performance requirement for reverse concatenation PS is a post-invDM BER of $<10^{-15}$, one needs to account for a potential BER increase due to the invDM operation. We will discuss this \sout{more} \uline{further} in Sec.~\ref{sec:perfmon}.

\subsection{Block performance monitoring}

In many modern communication systems, higher-layer protocols apply packet-oriented transmission, where the whole packet is discarded if any bit therein is received in error. BER metrics such as post-FEC BER or post-invDM BER might not be well suited to characterize the performance of such systems. Instead, systems using OTN framing are evaluated based on the \uline{\emph{background block error rate}} (BBER) and the \uline{\emph{severely errored second rate}}  (SESR) \cite{ITU-T_G.709,ITU-T_G.8201}. \uline{A severely errored second is a second with over 30 \% block errors.} The BBER is equal to the number of erroneous blocks normalized by the total number of transmitted blocks \sout{except for severely errored second (see Sec. II about the block size of OTUCn), and the} \uline{, disregarding those appearing during a severely errored second (the block size of OTUCn is given in Sec.~\ref{sec:system}). The} SESR is the probability of \sout{having a fraction of block errors at least the threshold, e.g., 30\%, 15\%, during one second} \uline{a severely errored second, and used as a long-term performance monitoring metric.} As BBER is a more basic performance metric, we will not consider SESR in this paper.

The BBER requirement depends on the link conditions, but a typical value is, e.g., around $10^{-7}$. For reverse concatenation PS, a large $N_{\text{s}}$ can cause a long error burst from the residual error after FEC decoding, although a large $N_{\text{s}}$ leads to smaller DM rate loss \sout{$\Delta$}\uline{$R_{\text{loss}}$}. As the block performance monitoring is critical for system design, we will simulate and analyze its behavior in the next section. Note that higher layer packets have smaller or larger sizes compared with the block size of OTUCn, so both the post-invDM BER and the BBER must be considered. In addition, we will consider the FEC frame error rate (FER), which is the probability of an erroneous FEC frame, or codeword, after FEC decoding.

\section{Implementation issues}
\label{sec:impl}

The state-of-the-art performance of PAS DM is provided by a coding scheme for $m$-out-of-$n$ codes \cite{ramabadran_1990_tcom} or CCDM \cite{schulte_2016_tit}. With this technique, a rate loss \sout{$\Delta$}\uline{$R_{\text{loss}}$} of almost zero is possible, provided that the DM word-length is large enough (typically $N_{\text{s}} \sim 10^3-10^4$). However, this requires a prohibitively complex hardware implementation. Recently, several attempts have been made to find a better balance between performance and complexity \cite{cho_2016_ecoc,pikus_2017_comlet,bocherer_2017_arxiv_spg,bocherer_2017_ecoc,yoshida_2017_ecoc_spg,yoshida_2018_ofc_spg,steiner_2018_ciss,fehenberger_2018_tcom,gultekin_2017_pimrc,gultekin_2018_isit,gultekin_2018_sitb,schulte_2018_arxiv}.
In \cite{cho_2016_ecoc,yoshida_2017_ecoc_spg}, fixed-length to fixed-length conversion was used, which is simple enough to implement, but the performance is limited.
In \cite{bocherer_2017_ecoc}, a fixed-length to variable-length conversion scheme with small address size look-up tables (LUT) was considered, showing a bounded conversion speed variation. However, this scheme can cause 
a large error burst after invDM 
if there is an error before the invDM.
A small LUT for a fixed-length to variable-length conversion was combined with a periodical uniformalization process in \cite{yoshida_2018_ofc_spg} to act as a fixed-length to fixed-length conversion. However, practical implementations of the fixed-length to variable-length conversion requires deeply sequential processing, and a large amount of parallelization (i.e., a large chip area) is required to realize the high throughput.
In \cite{pikus_2017_comlet,bocherer_2017_arxiv_spg,steiner_2018_ciss}, separate DM is applied to each modulation bit level for efficient implementation, and bit-wise DM like \cite{ramabadran_1990_tcom,yoshida_2017_ecoc_spg,yoshida_2018_ofc_spg} can be combined with it. \sout{However, the performance is not as good as the symbol-wise DM in the CCDM.} \uline{Its performance mainly depends on the bit-wise DM, and it is also useful for approaching the water-filling capacity on transmission over parallel channels having different SNRs \cite{bocherer_2017_arxiv_spg,che_2018,steiner_2018_ciss}.}
\uline{Other approaches to improve the performance with limited DM word length include partitioning and combining short constant-composition DM words \cite{fehenberger_2018_tcom}, enumerative sphere shaping and its approximations \cite{gultekin_2017_pimrc,gultekin_2018_isit,gultekin_2018_sitb}, and shell mapping \cite{schulte_2018_arxiv}. Their implementability on high-throughput signal processing still needs to be investigated.}

An FEC coding scheme has another drawback due to the throughput increase from reverse concatenation PS. Consider two cases, nonshaped BICM and reverse concatenation PS, assuming the same client \uline{(pre-DM) bit}rates, same FEC code rates $R_{\text{c}}$, and symbol rates. To distinguish the two cases, we introduce $m_{\text{u}}$ and $m_{\text{s}}$ as the number of modulation bits $m$ for the nonshaped and shaped cases, resp. The shaped signal requires $m_{\text{s}}/m_{\text{u}}$ times larger FEC throughput than the nonshaped one, because the bit rate is increased by the DM. Then the FEC circuit size and the power consumption will be increased by the ratio $m_{\text{s}}/m_{\text{u}}$ for the shaped case. Note that there is an FEC code rate restriction of $R_{\text{c}} \ge (m_{\text{s}}-1)/m_{\text{s}}$ in the case of PAS to \sout{put}\uline{assign} the uniformly distributed FEC parity bits to sign-bit positions without degrading the PMF of $P_{|X|}(|x|)$. In a generalization of the PAS, the parity bits can be placed outside the sign-bit, which relaxes this constraint \cite{yoshida_2017_ecoc_spg}.

\section{Proposed hierarchical DM (HiDM)}
\label{sec:prop}

\begin{figure}[tb]
	\begin{center}
		\setlength{\unitlength}{.6mm} %
		\scriptsize
		\vspace{-0.1cm}
		\includegraphics[scale=0.48]{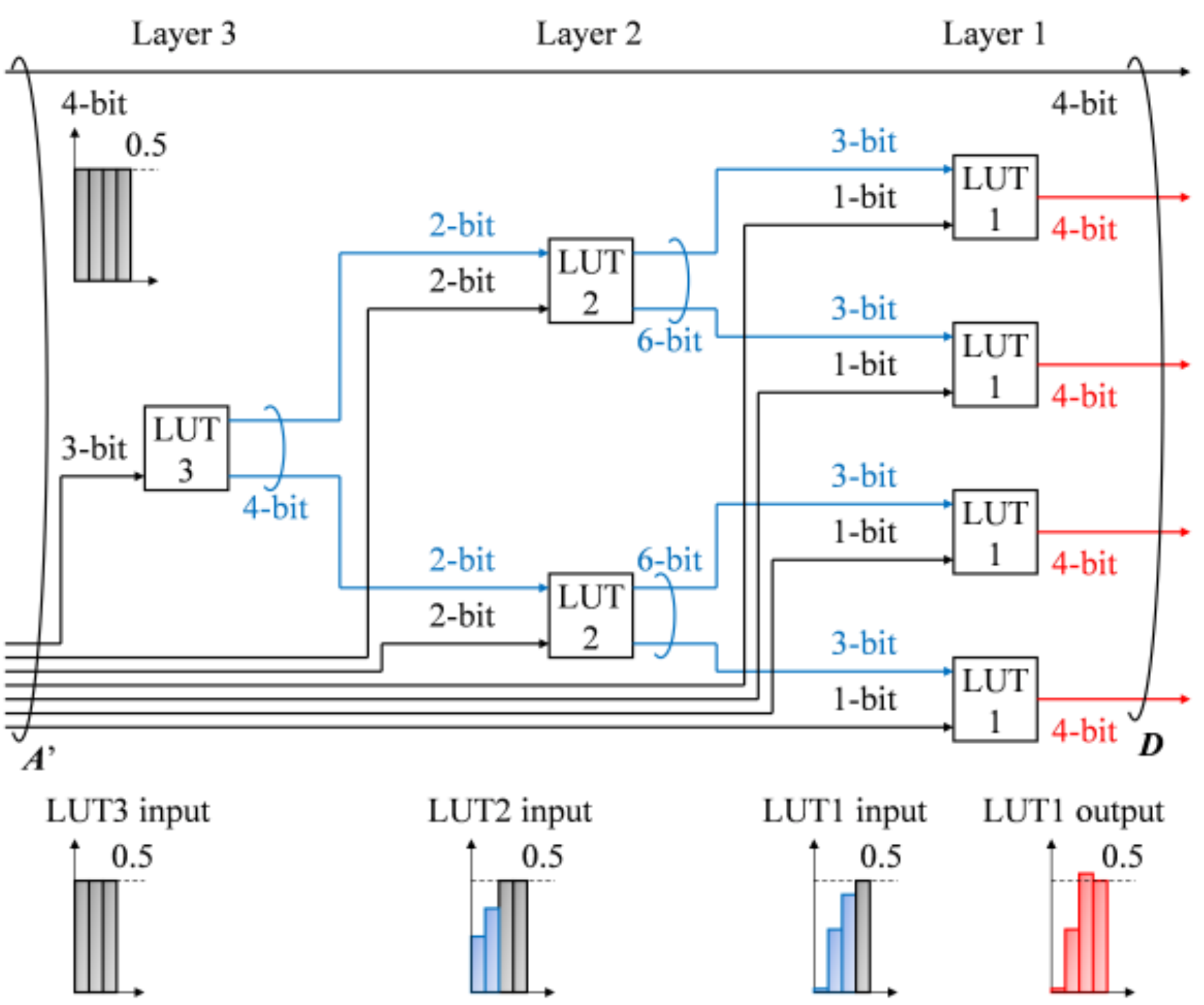} \\
		\vspace{-0.1cm}
		\caption{Example of the proposed hierarchical DM (HiDM), where $\uline{N_\text{u}=15}$ uniform input bits $\boldsymbol{A}'$ are converted into $\uline{m N_\text{s}=20}$ output bits $\boldsymbol{D}$ (4 uniform and 16 shaped bits). \uline{The non-shaped (uniform) bits will be used as sign bits of the 32-PAM symbols with $N_\text{s}=4$. The bar charts illustrate} \sout{bit}\uline{probabilities of being `1' for the respective bits.}}
		\label{fig:dm_example}
	\end{center}
\end{figure}

\begin{table}
	\caption{\uline{Bits ($\boldsymbol{B}'$) to symbol's absolute amplitude ($|X|$) conversion for 32-PAM.}}
	\label{tab:sym_mapg}
	\vspace{-0.4cm}
	\begin{center}
	\begin{tabular}{|cc|cc|cc|cc|}
		\hline
		$\boldsymbol{B}'$ & $|X|$ & $\boldsymbol{B}'$ & $|X|$ & $\boldsymbol{B}'$ & $|X|$ & $\boldsymbol{B}'$ & $|X|$ \\
		\hline\hline
		0000 & 1 & 0110 & 9 & 1100 & 17 & 1010 & 25 \\
		0001 & 3 & 0111 & 11 & 1101 & 19 & 1011 & 27 \\
		0011 & 5 & 0101 & 13 & 1111 & 21 & 1001 & 29 \\
		0010 & 7 & 0100 & 15 & 1110 & 23 & 1000 & 31 \\
		\hline
	\end{tabular}
	\end{center}
\end{table}

\begin{table*}
	\caption{The contents of the LUTs shown in Fig.~\ref{fig:dm_example}, including expected symbol energy and PMF.}
	\label{tab:LUTcontents}
	\vspace{-0.4cm}
	\begin{center}
	\setlength{\tabcolsep}{.6em}  
	\renewcommand\arraystretch{1.1}  
	\begin{tabular}{|cccc|cccc|cccc|}
		\hline
		\multicolumn{4}{|c|}{\sout{DM3} \uline{LUT3}} & \multicolumn{4}{c}{\sout{DM2} \uline{LUT2}} & \multicolumn{4}{|c|}{\sout{DM1} \uline{LUT1}} \\
		input & output & $\mathbb{E}[|X|^2]$ & PMF & 
		input & output & $\mathbb{E}[|X|^2]$ & PMF &
		input & output & $|X|$ & PMF \\
		\hline\hline
		000 & {\color{blue}00 00} & 21 & $1/8$ & {\color{blue}00} 00 & {\color{blue}000 000} & 5   & $1/8$  & {\color{blue}000} 0 & {\color{red}0000} & 1 & $23/128$ \\
		001 & {\color{blue}00 01} & 41 & $1/8$ & {\color{blue}00} 01 & {\color{blue}000 001} & 21  & $1/8$  & {\color{blue}000} 1 & {\color{red}0001} & 3 & $23/128$ \\
		010 & {\color{blue}01 00} & 41 & $1/8$ & {\color{blue}00} 10 & {\color{blue}001 000} & 21  & $1/8$  & {\color{blue}001} 0 & {\color{red}0011} & 5 & $11/64$ \\
		011 & {\color{blue}01 01} & 61 & $1/8$ & {\color{blue}00} 11 & {\color{blue}001 001} & 37  & $1/8$  & {\color{blue}001} 1 & {\color{red}0010} & 7 & $11/64$ \\
		100 & {\color{blue}00 10} & 63 & $1/8$ & {\color{blue}01} 00 & {\color{blue}000 010} & 53  & $1/16$ & {\color{blue}010} 0 & {\color{red}0110} & 9 & $3/32$ \\
		101 & {\color{blue}10 00} & 63 & $1/8$ & {\color{blue}01} 01 & {\color{blue}010 000} & 53  & $1/16$ & {\color{blue}010} 1 & {\color{red}0111} & 11 & $3/32$ \\
		110 & {\color{blue}00 11} & 83 & $1/8$ & {\color{blue}01} 10 & {\color{blue}001 010} & 69  & $1/16$ & {\color{blue}011} 0 & {\color{red}0101} & 13 & $3/64$ \\
		111 & {\color{blue}11 00} & 83 & $1/8$ & {\color{blue}01} 11 & {\color{blue}010 001} & 69  & $1/16$ & {\color{blue}011} 1 & {\color{red}0100} & 15 & $3/64$ \\
        	      &         &     &         & {\color{blue}10} 00 & {\color{blue}000 011} & 101 & $1/32$ & {\color{blue}100} 0 & {\color{red}1100} & 17 & $1/128$ \\
        	      &         &     &         & {\color{blue}10} 01 & {\color{blue}011 000} & 101 & $1/32$ & {\color{blue}100} 1 & {\color{red}1101} & 19 & $1/128$ \\
        	      &         &     &         & {\color{blue}10} 10 & {\color{blue}010 010} & 101 & $1/32$ & {\color{blue}101} 0 & {\color{red}1111} & 21 & 0 \\
        	      &         &     &         & {\color{blue}10} 11 & {\color{blue}001 011} & 117 & $1/32$ & {\color{blue}101} 1 & {\color{red}1110} & 23 & 0 \\
                      &         &     &         & {\color{blue}11} 00 & {\color{blue}011 001} & 117 & $1/32$ & {\color{blue}110} 0 & {\color{red}1010} & 25 & 0 \\
        	      &         &     &         & {\color{blue}11} 01 & {\color{blue}010 011} & 149 & $1/32$ & {\color{blue}110} 1 & {\color{red}1011} & 27 & 0 \\
        	      &         &     &         & {\color{blue}11} 10 & {\color{blue}011 010} & 149 & $1/32$ & {\color{blue}111} 0 & {\color{red}1001} & 29 & 0 \\
        	      &         &     &         & {\color{blue}11} 11 & {\color{blue}000 100} & 165 & $1/32$ & {\color{blue}111} 1 & {\color{red}1000} & 31 & 0 \\
		\hline
	\end{tabular}
	\end{center}
\end{table*}

Here we propose \sout{a hierarchical DM (HiDM) for efficient hardware implementation, which was briefly introduced in [22].} \uline{, and describe for the first time in detail, a \emph{hierarchical DM} (HiDM) with reasonable complexity, thus compatible with practical hardware limitations. It was briefly introduced in} \cite{yoshida_2018_ecoc}. 

\uline{Our aim is not to generate an arbitrary distribution of the transmitted symbols, but to approximate the Maxwell--Boltzmann (MB), or quantized Gaussian, distribution, which is the most commonly considered distribution in probabilistic shaping research. The MB distribution with parameter $\lambda$ is given by the PMF
\begin{IEEEeqnarray}{rCL}
\label{eq:MB}
P_X^{\text{MB}} (x) = \frac{\text{exp} (-\lambda x^2)}{\sum_{\xi \in \mathcal{X}} \text{exp} (-\lambda \xi^2)}
\end{IEEEeqnarray}
for any $x \in \mathcal{X}$. For a given base constellation and average symbol energy, the MB distribution maximizes the entropy \cite{kschischang_1993_tit}. Furthermore, it approximates well the distribution that maximizes the mutual information over the Gaussian channel under an average energy constraint \cite{raphaeli_2004_tcom}. Conversely, if the information bit rate and base constellation are fixed, minimizing the average symbol energy results in a distribution similar to MB, as will be exemplified in Sec.~\ref{sec:sim}.}

\uline{In principle, a DM with the best performance at a finite DM word length could be realized by a single look-up table (LUT), whose average symbol energy corresponding to the list of output bits $\boldsymbol{D}$ is minimized. However, when the DM word length $N_{\text{s}}$ is large such as 20, the LUTs for DM and invDM are prohibitively large to be implemented in hardware. Thus we will demonstrate how to use small- to medium-size LUTs and combine them hierarchically to construct a larger set of output bit sequences than what each LUT can provide.}

The input to the DM in Fig.~\ref{fig:system} is the bit sequence $\boldsymbol{A}' \in \mathcal{B}^{N_{\text{u}}}$. Of those bits, $N_{\text{u}}^{\text{sb}}$ are converted by the shaping. The number of shaped bit levels \uline{per QAM symbol} is $m^{\text{sb}}$, \sout{and the number of shaped bits is hence $m^{\text{sb}} N_{\text{s}}$, which is a subset of the DM output bits $\boldsymbol{D} \in \mathcal{B}^{m N_{\text{s}}}$} \uline{which means that $m^{\text{sb}} N_{\text{s}}/ 2$ of the $m N_{\text{s}} /2$ DM output bits $\boldsymbol{D}$ are shaped.}
The parameters $m$, $m^{\text{sb}}$, $N_{\text{u}}$, $N_{\text{u}}^{\text{sb}}$, $N_{\text{s}}$ \uline{are dependent. The number of unshaped bits per DM word must be the number of FEC encoder input bits $R_{\text{c}}mN_{\text{s}}/2$ subtracted by the number of shaped bits $m^{\text{sb}}N_{\text{s}}/2$, i.e,}\sout{have a relation with FEC code rate $R_{\text{c}}$ as}
\begin{IEEEeqnarray}{rCL}\label{eq:unshaped}
	\uline{N_{\text{u}} - N_{\text{u}}^{\text{sb}} = \left( R_{\text{c}} m  - m^{\text{sb}} \right) N_{\text{s}} / 2.}
\end{IEEEeqnarray}
Next we will explain how $N_{\text{u}}^{\text{sb}}$ uniform input bits can be converted into $m^{\text{sb}} N_{\text{s}} /2$ nonuniform \sout{(probabilistically shaped)} bits \uline{that generate probabilistically shaped output symbols.}

\subsection{Principle of operation}
\label{sec:principle}
A small example of the proposed HiDM for generating a sequence of four PS-32-PAM symbols is shown in Fig.~\ref{fig:dm_example}, where \sout{$m=5$}\uline{$m/2=5$}, the shaped bit levels \uline{for a PAM symbol} $\boldsymbol{i}_{\text{sb}} = (2,3,4,5)$, $m^{\text{sb}}=$ \uline{$2$} $|\boldsymbol{i}_{\text{sb}}|=$\sout{$4$}\uline{$8$}, $N_{\text{u}}^{\text{sb}} = 11$, and $N_{\text{s}} = 4$ (i.e., four PAM symbols \uline{or two QAM symbols}). For simplicity, we assume here an FEC code rate of $R_{\text{c}} = 1$. Hence, the DM has 15 uniform input bits and 20 output bits.
Four of the input bits remain untouched by the DM, while 11 input bits are converted into 16 shaped bits.
\uline{Tab.~\ref{tab:sym_mapg} shows the assumed bit-to-symbol mapping table for 32-PAM.}
The DM consists of three small \sout{DMs, called DM1, DM2, and DM3,} \uline{LUTs, called LUT1, LUT2, and LUT3}, hierarchically organized in a tree-like structure with three layers. Each layer comprise\uline{s} one or more \sout{DMs} \uline{LUTs}, which are the same within each layer. The LUT contents including the expected one-dimensional \sout{power} \uline{energy} $\mathbb{E}[|X|^2]$ and probabilities of each small-DM word, PMF, are shown in Tab.~\ref{tab:LUTcontents}.

At first, three uniformly distributed information bits are input to \sout{DM3} \uline{LUT3} and converted into four output bits, which will act as constraint bits in the connected lower-layer DMs. The four constraint bits are separated into two lines of two constraint bits, each having mark ratios\footnote{A ``mark'' is a bit equal to `1', and the probability for that bit being `1' is commonly referred to as ``mark ratio''.} of 0.25 and 0.375, which together with two uniformly distributed information bits are fed into each \sout{DM2} \uline{LUT2. The behavior of the resulting bit probability statistics, i.e., mark ratios, are shown by the bar charts in Fig.~\ref{fig:dm_example}.} The constraint bits are chosen so that the expected \uline{symbol} energy becomes small. \sout{This corresponds to that each list in Tab.~\ref{tab:LUTcontents} is sorted by the expected energy.} \uline{LUT2} converts four input bits into six output bits, which are again separated into two lines and treated as constraints bits in the next layer. \sout{Complementing} \uline{Inputting} each set of three constraint bits having \sout{expected} mark ratios of 0.016, 0.281, and 0.438 with one uniformly distributed information bit, each \sout{DM1} \uline{LUT1} generates four output bits having \sout{expected} mark ratios of 0.016, 0.300, 0.531, and 0.5 \uline{shown by red bar charts}. Note that the output bits from \sout{DM3, DM2, or DM1} \uline{the LUTs} are dependent. The output bits from \sout{DM1} \uline{LUT1} will now determine the absolute amplitudes $|\boldsymbol{X}|$ of the symbol sequence $\boldsymbol{X}$. In this ex\uline{a}mple, the \sout{probabilities of $|X|$ becomes 0.445, 0.445, 0.437, 0.437, 0.320, 0.320, 0.207, 0.207, 0.055, 0.055, 0, 0, 0, 0, 0, 0 for $|X| = 1, 3, \ldots , 31$.} \uline{PMF of $|X|$ becomes nonuniform and one-sided MB-like, as shown in the last column of Tab.~\ref{tab:LUTcontents}. The generated DM words are not constant composition.}

The expected \sout{power} \uline{energy} in Tab.~\ref{tab:LUTcontents} is derived from the expected \sout{powers} \uline{energies} in the lower layer\sout{,}\uline{s.}  \uline{E.g., the LUT2 entry with output bits `000 001' generates input sequences `000u' and `001u' to layer 1, where `u' denotes a uniform choice of `0' or `1'. The average energies of these sequences are $(1^2+3^2)/2=5$ and $(5^2+7^2)/2=37$, resp., and hence $\mathbb{E}[|X|^2] = (5+37)/2 = 21$ for `000 001' in LUT2, as shown on the second row of Tab.~\ref{tab:LUTcontents}.}
\sout{e.g., $\mathbb{E}[|X|^2]$ for output bits "000 001" (the second small DM word) in DM2 is the average energy for a bit input to DM1 "000d" (i.e., $(1^2+3^2)/2=5$) and "001d" (i.e., $(5^2+7^2)/2=37$), where "d" denotes don't care of "0" or "1". Then $\mathbb{E}[|X|^2]$ of $(5+37)/2 = 21$ is obtained, as shown in the second word in DM2.}
Each list is sorted from lower to \sout{upper} \uline{higher} \sout{powers} \uline{energy}, and lower \sout{power ones are} \sout{energy will be} chosen \uline{more} frequently by the constraints from \sout{upper} \uline{higher} layers. 

\uline{From these values, the} \sout{The} information rate of the PAM symbol is $15/20\times 5 =3.75$, and the entropy is $3.93$, and the rate loss \sout{$\Delta$}\uline{$R_{\text{loss}}$} becomes \sout{$0.177$}\uline{$0.18$} \sout{in one} \uline{bits per} dimension. When the five output bits are mapped to a Gray-coded 32-PAM symbol, the expected energy per one-dimensional symbol is 57. The \uline{\emph{constellation gain}, defined as \cite[Eq.~(8)]{kschischang_1993_tit}
\begin{IEEEeqnarray}{rCL}
\label{eq:CG}
G = \frac{d_{\text{min}}^2 (2^{\beta}-1)}{6E},
\end{IEEEeqnarray}
where $d_{\text{min}}$, $\beta$, and $E$ denote minimum Euclidean distance, spectral efficiency (bpcu) at an FEC code rate of 1, and average symbol energy, resp.,
}
will be 0.22~dB.


\sout{The proposed hierarchical architecture constrain the output bits from the connected lower-layer LUTs. The upper-layer LUT is designed so that its output roughly matches the desired bit distribution by constraining the subset of the output bits at the lower layers, and the lower-layer DMs gradually refine the distribution for better agreement with the target.}

Additional flexibility and finer granularity can be obtained by increasing the number of layers, the number of \sout{DMs} \uline{LUTs} connected between layers, the number of input information bits to each layer's \sout{DM} \uline{LUT}, and the number of constraint bits in each layer, which gives better \sout{control}\uline{adaptation} of the \sout{output signal distribution} \uline{PMF $P_X (x)$ to an approximated MB distribution} and more significant performance improvements.
\sout{Each DM function is one-to-one correspondence, so it is reversible at the invDM.}
\uline{Each LUT and thus the whole DM function are one-to-one mappings, so that they are reversible for the invDM. }
\uline{All constraint bits are used for the selection of symbols to minimize the expected symbol energy. The recipes for designing the LUTs will be discussed next in  Sec.~\ref{sec:recipe}.}

\subsection{Recipe for LUT-tree construction and LUT content design}
\label{sec:recipe}
\uline{In this section, we describe how to design the HiDM, to realize flexible rates and/or maximization of the constellation gain at given input/output rates. We denote by $L$ the number of layers and by $T_\ell$ the number of LUTs in layer $\ell=1,\ldots, L$, where $T_L = 1$. The number of input bits to each LUT on layer $\ell$ is denoted by $v_\ell = r_\ell+s_\ell$, where $r_\ell$ bits come from an LUT on layer $\ell+1$ and $s_\ell$ bits are take directly from the DM input. Each LUT on layer $\ell+1$ is connected to $t_\ell=T_\ell/T_{\ell+1}$ different LUTs on layer $\ell$, and hence the number of LUT output bits is $u_{\ell+1} = t_\ell r_\ell$. The total number of shaped DM output bits is $T_1 u_1$, corresponding to $T_1 u_1/(m/2-1)$ PAM symbols, and the corresponding number of DM input bits is $N_\text{u}^\text{sb} = \sum_\ell T_\ell s_\ell$. The number of unshaped bits passing untouched through the HiDM is given by \eqref{eq:unshaped}.}

\uline{Using this notation, the HiDM LUT tree is designed by repeating the following steps for $\ell=1,2,\ldots$:}
\begin{enumerate}
	\renewcommand{\labelenumi}{\arabic{enumi})}
	\item \uline{Select the number of output bits $u_\ell$ of each LUT in layer $\ell$. The number of output bits should not be too large, like 20 or above, since it would require a big memory size in the invDM. Suitable values are in the range from 2 to 14 bits, e.g., $(u_1,u_2,u_3) = (4,6,4)$ bits in Fig.~\ref{fig:dm_example}.}
	\item \uline{Select the number of input bits $s_\ell$
that are taken from the DM input, observing that the rate will not be so flexible if $s_{\ell}/u_{\ell}$ is small.
Also select the number of input bits $r_\ell$ that come from LUTs in layer $\ell+1$.
The higher-layer LUT storage size will be big if this value is too high. In Fig.~\ref{fig:dm_example}, $(s_1,s_2,s_3) = (1,2,3)$ and $(r_1,r_2,r_3) = (3,2,0)$.}
	\item \uline{Select the number of LUTs $t_\ell$ in layer $\ell$ connected to a single LUT in layer $\ell+1$.
}
	\item \uline{Terminate when number of output PAM symbols $T_1 u_1/(m/2-1)$ is sufficient, where $T_1 = t_1 t_2 \cdots t_\ell$. Let $L=\ell$.}
\end{enumerate}

\uline{ Once the LUT-tree configuration is established, the contents in each LUT is determined by the following sorting method.}
\uline{The contents of the LUTs are another critical matter for HiDM, so its basic concept is shown here, again using the small example shown in Tab.~\ref{tab:LUTcontents}.}
\begin{enumerate}
	\renewcommand{\labelenumi}{\arabic{enumi})}
	\item \uline{List the output DM words of each LUT in layer 1 by sorting them after increasing symbol energy, as shown in LUT1 in Tab.~\ref{tab:LUTcontents}. There are $2^{u_1}$ candidates, and the top $2^{v_1}$ words (that is, those having smallest symbol energy) are selected. 
	}
	\item \uline{For $\ell = 2,\ldots, L$, calculate the expected energies for all possible output words from the LUT in layer $\ell$, using the procedure described in Sec.~\ref{sec:principle}.
Sort these words by increasing energy, as shown in LUT2 in Tab.~\ref{tab:LUTcontents}. There are $2^{u_\ell}$ candidate words, of which the top $2^{v_\ell}$ are valid.}
\end{enumerate}

\uline{If we change the parameters $(r_1, s_1, u_2, s_3)$ from $(3,1,6,3)$ in Fig.~\ref{fig:dm_example} into $(2,2,4,4)$, the number of input bits and the DM information rate in the HiDM becomes 20 and 20/20 = 1, resp. We can also change $s_{\ell}$ to have a desired DM code rate with a granurality of $1/20$. When implementing a fully flexible-rate HiDM in hardware, the LUT architecture should support a DM information rate of 1, an external circuit should control the used/unused information bit input path, and the LUT contents should ignore the unused bits.}

\subsection{Complexity}
\label{sec:complexity}
The complexity of the required circuitry is dominated by the number of stored bits in the LUTs, which determines the area in the hardware implementation.
The number of stored bits
\sout{depends on the number of words (addresses), the number of output bits, and the number of LUTs}\uline{for all LUTs in layer $\ell$ is $T_\ell 2^{v_\ell}u_\ell$}.
In the example of Fig.~\ref{fig:dm_example}, the number of stored bits are \sout{$2^3 \times 4 \times 1 = 32$ for LUT3, $2^4 \times 6 \times 2 = 192$ for LUT2, and $2^4 \times 4 \times 4 = 256$ for LUT1}\uline{32, 192, and 256 in layers 3, 2, and 1, resp.}, thus in total 480. The corresponding LUTs on the invDM side have
\sout{$2^4 \times 3 \times 1 = 48$ stored bits in layer 3, $2^6 \times 4 \times 2 = 512$ in layer 2, and $2^4 \times 4 \times 4 = 256$ in layer 1, which totals 816 stored bits.}\uline{$T_\ell 2^{u_\ell}v_\ell$ stored bits, which in the example gives $48+512+256 = 816$ bits in total for the three layers.}
If we would instead use a single LUT, i.e., 11-bit input and 16-bit output on the DM side and conversely on the invDM side, the number of stored bits would be $2^{11} \times 16 = 32768$ and $2^{16} \times 11 = 720896$, resp. Thus, HiDM saves circuit resources by a factor of $68$ and $883$ in the DM and invDM, resp., already in this small-scale example. \uline{Note that the performance of HiDM is always inferior to a DM with a single LUT that can have the optimum list (lowest average symbol energy) for the DM words, but the gain in complexity and memory size can be considerable,}

\uline{HiDM} \sout{DM schemes} with \uline{DM} word length $N_{\text{s}} \approx 100$ or $1000$ can be realized by LUTs with reasonable sizes. 
The DM does thus not require complex operations such as integer additions or multiplications. The conversion process is purely fixed-length input and fixed-length output (there is no variable length part). Either bit-wise \cite{pikus_2017_comlet} or symbol-wise \cite{schulte_2016_tit} DMs are possible, and 
\sout{arbitrary output}\uline{MB-like} distributions can be approximated with high granularity.
\sout{This is not constant composition DM, but its PMF will converge to the target for sufficiently many DM words.} 
This DM achieves a high throughput because the architecture consists of a fully parallelized input/output configuration, as well as a bit-scramble selector and a permutation mapper \cite{yoshida_2017_ecoc_spg}. The hierarchical operation is fully pipelined, so a small number of instances is enough. 

In contrast, \uline{arithmetic coding based DM, e.g.,} a coding scheme for $m$-out-of-$n$ codes \cite{ramabadran_1990_tcom} \sout{such as} \uline{and} \sout{the} CCDM \cite{schulte_2016_tit} requires high-precision integer multiplications, which limits the throughput, and a large number of instances (at least $\sim N_{\text{s}}$) would be required\footnote{\label{footnote:instance}\uline{Here ``instance'' refers to the number of DM circuits that have to be implemented in parallel on the ASIC. While the latest CMOS logic circuitry can operate at hundreds of GHz, the throughput of optical communication links reaches hundreds of Gb/s to Tb/s. CCDM requires sequential operation, so one CCDM circuit can generate the absolute amplitude of one PAM symbol per clock cycle using a straight-forward implementation, although it depends on the logical circuit design. Then to generate $N_{\text{s}}$ PAM absolute amplitudes, $N_{\text{s}}$ clock-cycles are needed. Until completing one DM word, one cannot input information bits for the next DM word. Thus $N_{\text{s}}$ parallel CCDM circuits are necessary for real-time operation. 
When the clock speed is 500 MHz, 320 Gsymbol/s is achieved, which would be sufficient for the required throughput. On the contrary, HiDM outputs, e.g., 320 Gsymbol/s per HiDM circuit at a clock-cycle, so it requires just a few instances.}}. Then the equivalent DM word-length is in the order of $N_{\text{s}}^2$ or more, as will be further discussed in the next section. 

\subsection{Frame structure}

\begin{figure}[tb]
	\begin{center}
		\setlength{\unitlength}{.6mm} %
		\scriptsize
		\vspace{-0.1cm}
		\includegraphics[scale=0.45]{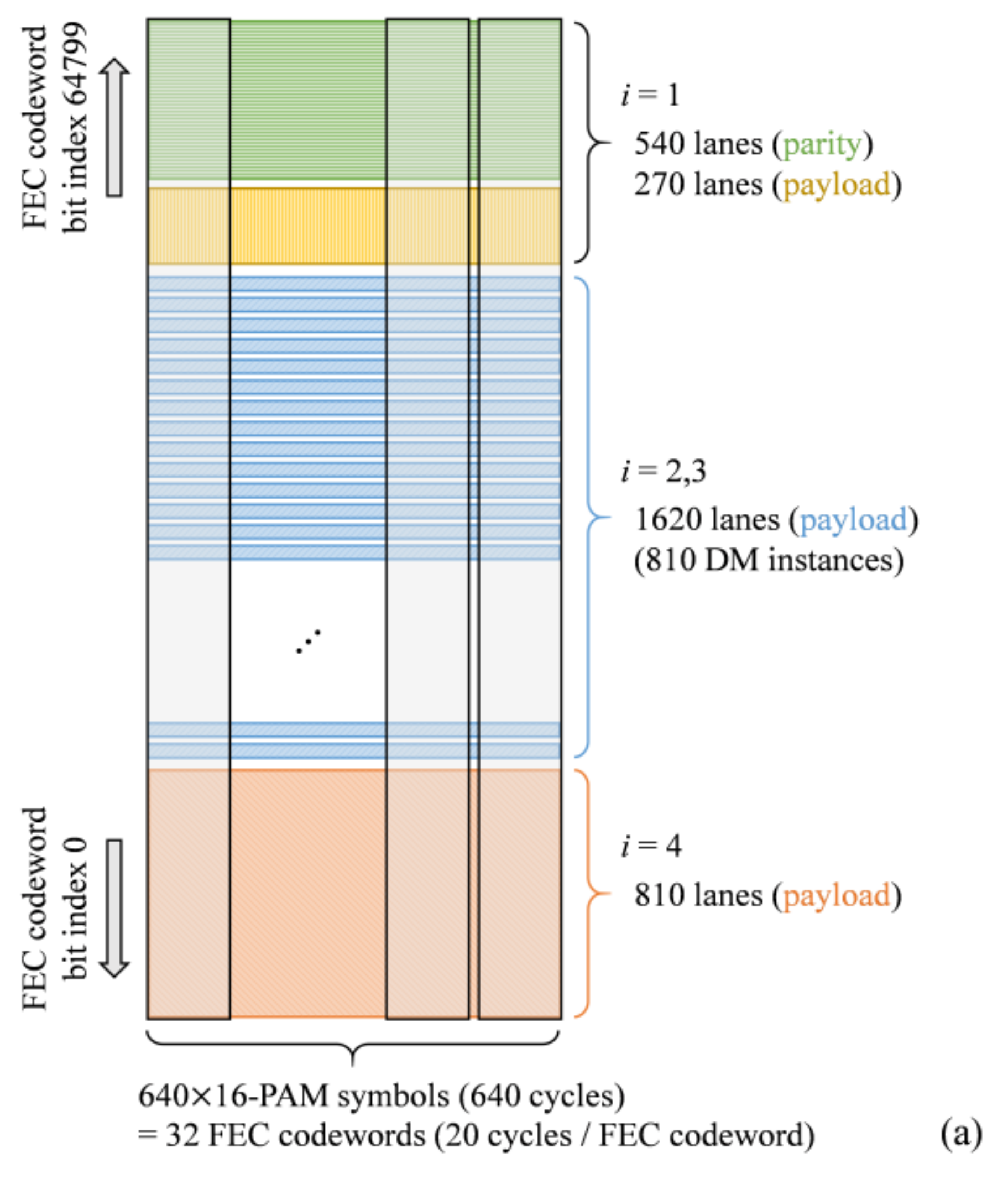} \\
		\includegraphics[scale=0.45]{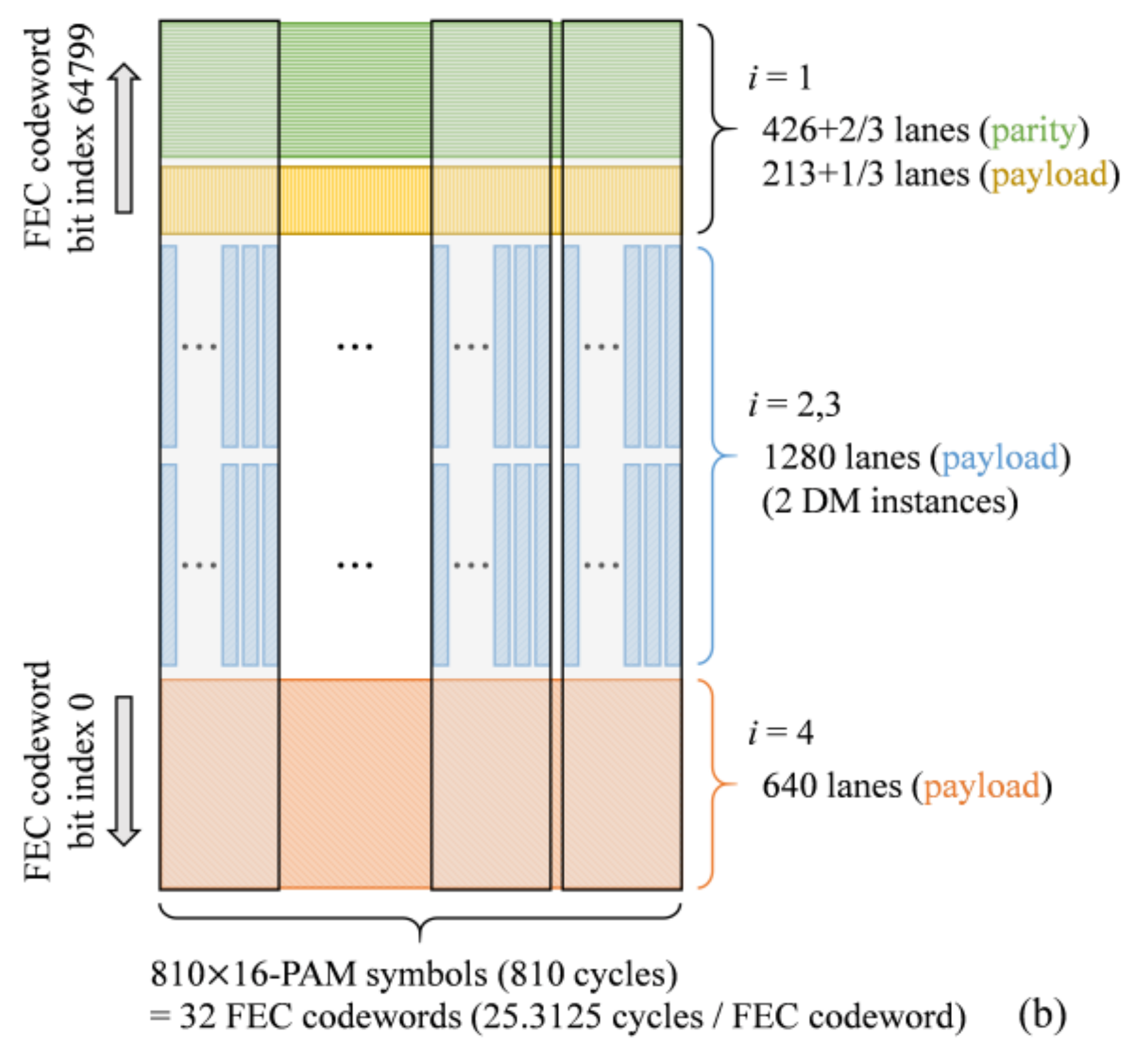} \\
		\vspace{-0.3cm}
		\caption{\sout{Line-side f}\uline{F}rame structure with (a) CCDM or with (b) HiDM. \uline{One blue (filled) rectangle shows one DM word.}}
		\label{fig:frame}
	\end{center}
\end{figure}

An example of a \sout{line-side} frame structure \uline{used for the transmission line (channel)} with the CCDM (a) and the proposed HiDM (b) are shown in Fig.~\ref{fig:frame}. 
\uline{The blue (filled) rectangles show one DM word. As discussed in Footnote \ref{footnote:instance}, while one CCDM circuit generates one symbol per  clock-cycle due to the sequential operation, HiDM generates many symbols per clock-cycle because of the parallel output interface. This leads to the difference in Figs.~\ref{fig:frame}(a) and (b)}%
\footnote{\uline{A HiDM-like frame structure for CCDM can be realized by rearrangement of the memory, but it will significantly increase the latency.}}\uline{.}
The FEC is assumed to be a DVB-S2 low-density parity-check code \cite{dvbs2} with codeword length $n_\text{c} = 64800$ and code rate $R_{\text{c}} = 5/6$. The modulation format is Gray-coded 16-PAM. The most significant ($i=1$, sign-bit) and the least significant ($i=4$) bit-levels for the 16-PAM symbols are not shaped to simplify the implementation. 
Thus, $m^{\text{sb}}\uline{/2}=2$ and the shaped bit levels are $\boldsymbol{i}_{\text{sb}}=\{2,3\}$.
The number of input bits $N_{\text{u}}^{\text{sb}}$ is 1014 or 507, the number of output \uline{PAM} symbols $N_{\text{s}}$ is 640 or 320, and the number of output bits $m^{\text{sb}} N_{\text{s}}$\uline{$/2$} is 1280 or 640 for CCDM or HiDM, resp., in this example. 
The HiDM is designed using seven layers. In layer 1, there are 64 small \sout{DMs (DM1)} \uline{LUTs}, each having $u_1=10$ output bits. Thus, the total number output bits from the proposed DM is $640$, which equals $m^{\text{sb}} N_{\text{s}}\uline{/2}$.
\uline{The required storage sizes for HiDM with these parameters and two instances are manageable, with values of $3.7 \times 10^6$ for DM and $6.8 \times 10^6$ for invDM, while that for two instances using a single LUT would be prohibitively large, i.e., $2\cdot 2^{507} \cdot 640 = 5.4 \times 10^{155}$ for DM and $2 \cdot 2^{640} \cdot 507 = 4.6 \times 10^{195}$ for invDM. Thus, HiDM reduces the total storage size by more than $10^{194}$ times, though there is a performance drawback compared to using a single LUT.} 

The information rate \uline{$R$} \sout{($R-\Delta$)} is 5.83 \sout{bit/channel use}\uline{bpcu}, which is equivalent to 128-QAM with $R_{\text{c}}=5/6$. The equivalent word-length for CCDM and HiDM are 518400 ($>N_{\text{s}}^2$) and 320, resp. While 810 DM words are equally mapped to 32 FEC codewords in parallel for the CCDM, 50 or 51 DM words are included in an FEC codeword sequentially for the HiDM.
The bit indices 0--53999 of the FEC codeword correspond to the payload bits and indices 54000--64799 to the parity bits. The less significant bits are placed at lower bit indices of the FEC codeword to balance the pre-FEC performance and the FEC decoding capability.

\section{Simulations}
\label{sec:sim}
To evaluate the post-FEC and post-invDM performance with reverse concatenation PS, we conducted numerical simulations of PS-256-QAM transmission over the Gaussian channel. The same combinations of FEC and PS were examined as in Sec.~\ref{sec:prop}, i.e., the DVB-S2 low-density parity-check code with $R_{\text{c}}=5/6$. The number of decoding iterations was 20 and more than 1600 FEC codewords were examined per simulation. The soft-demapping input and output interfaces were quantized with 7 and 4 bits, resp., which gives less than $0.1$~dB SNR penalty. Two DM schemes were implemented, HiDM and CCDM. For comparison we also simulated nonshaped BICM 128-QAM \uline{with a Gray-like labeling} \cite[Slide~10]{mahadevappa_2003}.%
\footnote{\uline{There is no perfect Gray code for cross-QAM constellations such as this 128-QAM, which leads a performance drawback.}}
Its most significant bit ($i=1$) and parts of the second significant bit ($i=2$\uline{, which are sign bits}) were occupied by the FEC parity bits, for the same information rate.\footnote{\uline{Performance comparisions between shaped and non-shaped signals raises the question of whether to use different base constellations or different FEC code rates. For ease of hardware implementation and flexible setting of spectral efficiencies, we chose the same FEC code rate but different base constellations in this paper.}}

\uline{Before evaluating the error rates, we summarize the statistics of the PMF $P_{|X|}(\cdot)$ in Tab.~\ref{tab:stat_spg} for CCDM, HiDM, and the MB distribution with $\lambda = 0.1373$. We also give the average symbol energy $E$, the two-dimensional entropy $2H(X)$, the spectral efficiency $\beta$ at $R_{\text{c}}=1$, the DM rate loss $R_{\text{loss}}$ (which equals $2H(X) - \beta$ in this case), and the constellation gain $G$. For CCDM, $\beta$ can be increased to $2\cdot (2+1015/640)$ at a target PMF of \eqref{eq:MB} with $\lambda = 0.0133$, but it was reduced to $2\cdot (2+1014/640)$ to fit to the frame. Its influence on $R_{\text{loss}}$ and $G$ is 0.004 bpcu and 0.01 dB, resp. HiDM shows 0.13 dB and 0.39 dB performance loss compared with CCDM and the MB distribution, resp. If the DM word length $N_{\text{s}}$ of CCDM is decreased from 640 to 320, then the constellation gain $G$ becomes almost the same as HiDM.}

\begin{table}
	\caption{\uline{Resulting statistics by shaping.}}
	\label{tab:stat_spg}
	\vspace{-0.4cm}
	\begin{center}
	\begin{tabular}{|c|cc|c|c|}
		\hline
		& \multicolumn{2}{|c|}{CCDM} & HiDM & MB \\
		\hline\hline
		$N_{\text{s}}$ (PAM-symbol) & 640 & 320 & 320 & -- \\
		$P_{|X|}(1)$ & 0.2484 & 0.2453 & 0.2376 & 0.2628 \\
		$P_{|X|}(3)$ & 0.2484 & 0.2453 & 0.2376 & 0.2355 \\
		$P_{|X|}(5)$ & 0.1625 & 0.1625 & 0.1684 & 0.1891 \\
		$P_{|X|}(7)$ & 0.1625 & 0.1625 & 0.1684 & 0.1360 \\
		$P_{|X|}(9)$ & 0.0695 & 0.0719 & 0.0757 & 0.0877 \\
		$P_{|X|}(11)$ & 0.0695 & 0.0719 & 0.0757 & 0.0506 \\
		$P_{|X|}(13)$ & 0.0195 & 0.0203 & 0.0183 & 0.0262 \\
		$P_{|X|}(15)$ & 0.0195 & 0.0203 & 0.0183 & 0.0121 \\ \hline
        	$E$ & 72.50 & 74.00 & 74.70 & 68.31 \\
        	$2H(X)$ (bpcu) & 7.214 & 7.242 & 7.252 & 7.169 \\
        	$\beta$ at $R_{\text{c}} = 1$ (bpcu) & 7.169 & 7.169 & 7.169 & 7.169 \\
        	$R_{\text{loss}}$ (bpcu) & 0.045 & 0.073 & 0.083 & 0 \\
		$G$ (dB) & 1.186 & 1.097 & 1.056 & 1.444 \\
		\hline
	\end{tabular}
	\end{center}
\end{table}

\begin{figure}[tb]
	\begin{center}
		\setlength{\unitlength}{.6mm} %
		\scriptsize
		\vspace{-0.1cm}
		\includegraphics[scale=0.48]{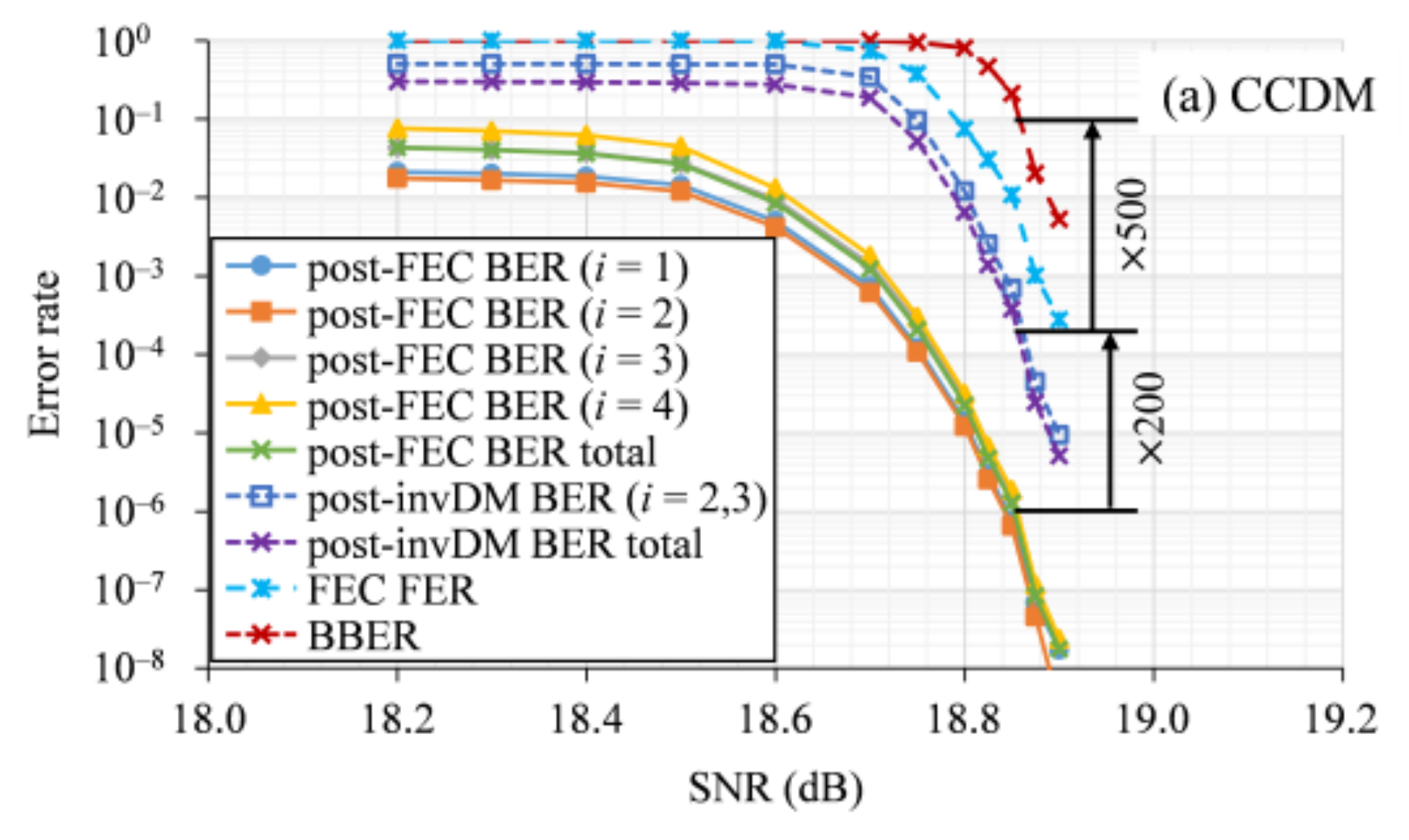} \\
		\includegraphics[scale=0.48]{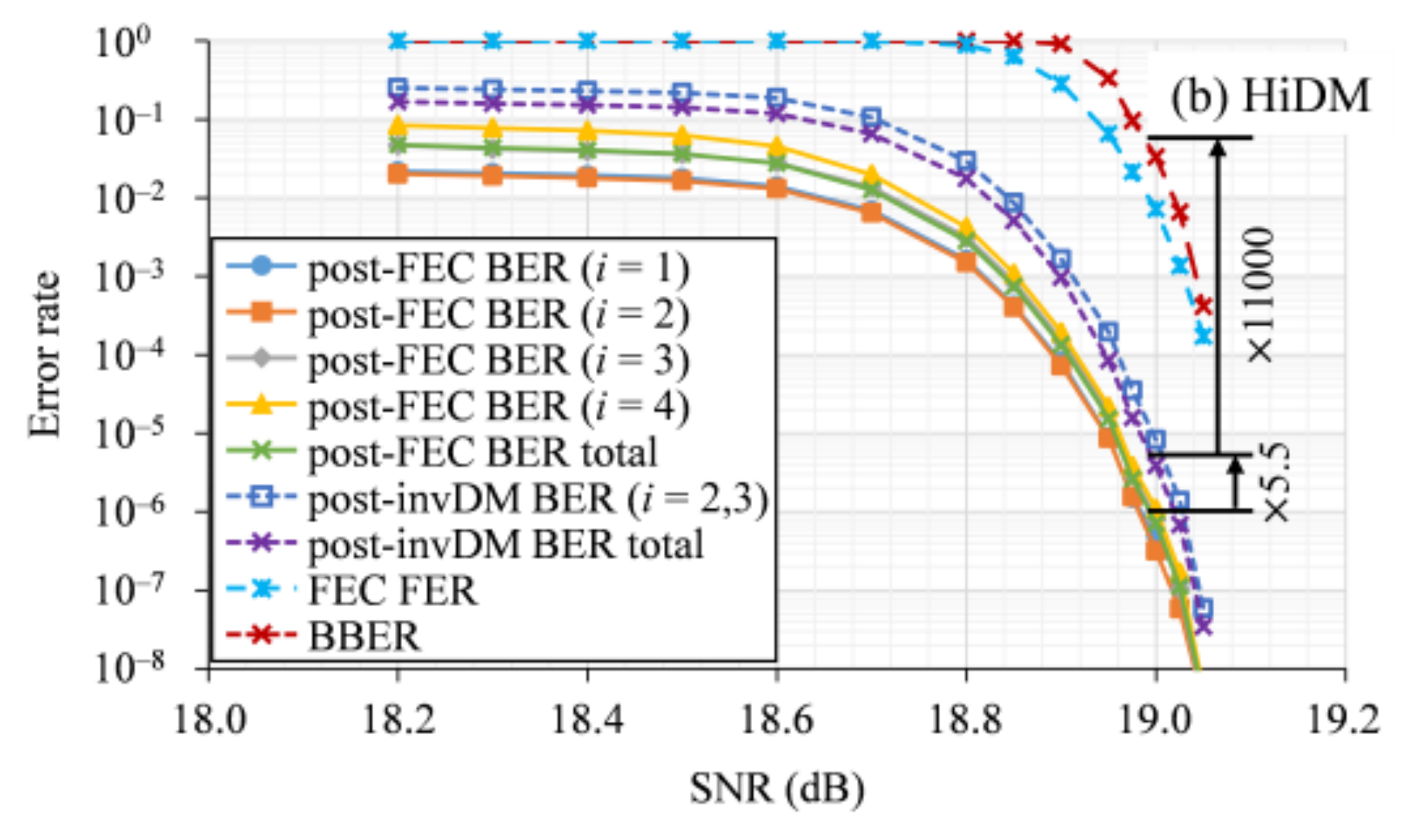} \\
		\includegraphics[scale=0.48]{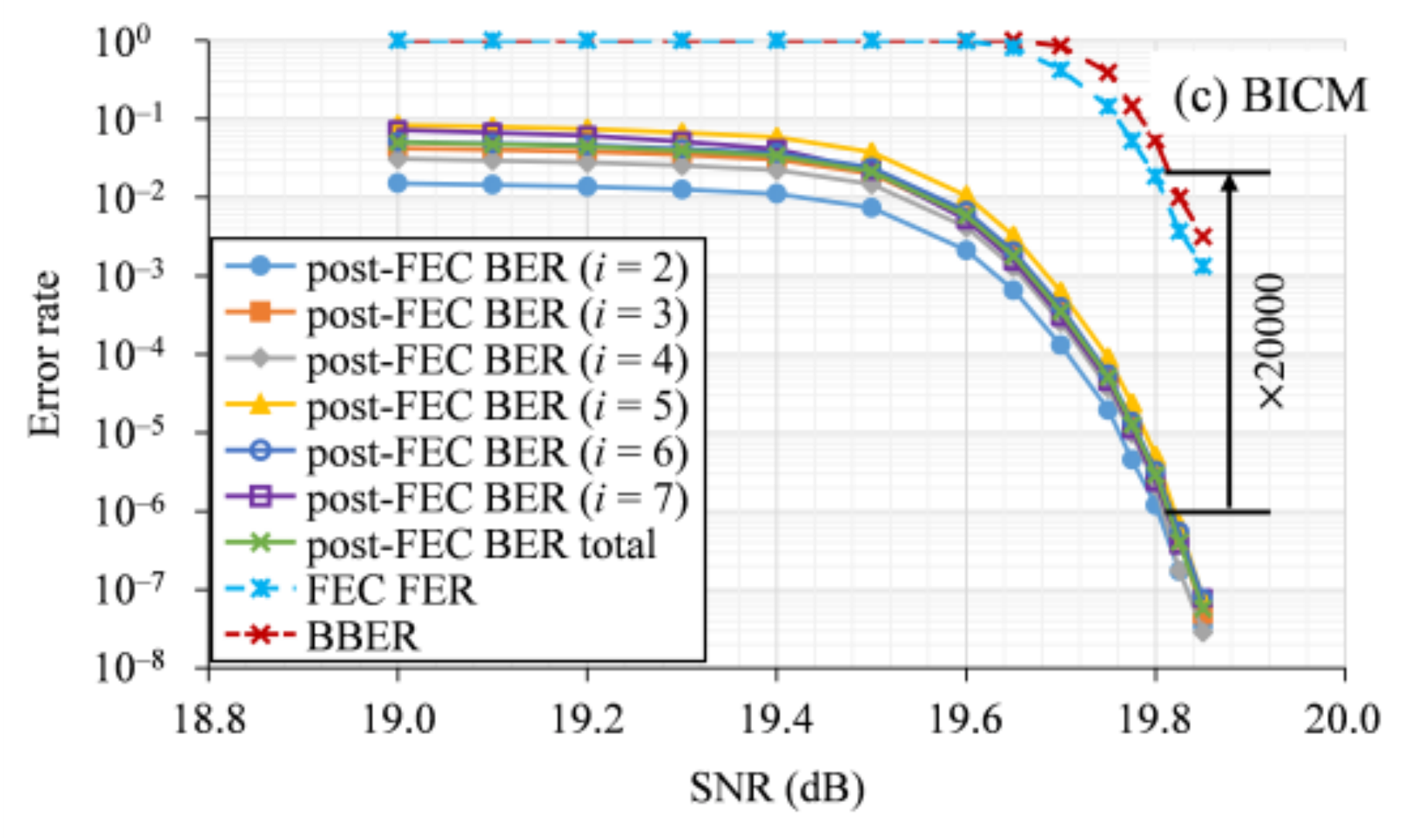} \\
		\vspace{-0.3cm}
		\caption{Simulation results of (a) CCDM-based PS-256-QAM, \uline{(b)} HiDM-based PS-256-QAM, and (c) BICM-based 128-QAM.}
		\label{fig:simres}
	\end{center}
\end{figure}

Fig.~\ref{fig:simres} shows the simulation results in terms of post-FEC BER, post-invDM BER, FER, and BBER as a function of the SNR. Here we assume the use of OTUC1 framing, so the block size is 130560 bits. The post-FEC BER at low BERs is similar among the bit levels, due to the bit level mapping in the FEC codeword space \cite{bocherer_2015_tcom}. CCDM shows $0.13$~dB lower required SNR than HiDM at a post-FEC BER of $10^{-6}$, and 128-QAM BICM performed further $0.8$~dB worse than HiDM. \uline{The difference in the required SNR of 0.13 dB between CCDM and HiDM agrees well with the difference in constellation gain $G$ shown in Tab.~\ref{tab:stat_spg}.}
\uline{If we decrease the DM word length $N_{\text{s}}$ to 320 for CCDM, the required SNR is expected to increase by around 0.1~dB, according to the difference in constellation gain $G$.}

The error rate increase by the invDM for the reverse concatenation PS is characterized by the post-invDM BER $\mathcal{E}_{\text{post-iDM}}$ to post-FEC BER $\mathcal{E}_{\text{post-FEC}}$ ratio,
\begin{IEEEeqnarray}{rCL}
	\label{eq:rate_BER}
	r_{\mathcal{E} 1} &=& \frac{\mathcal{E}_{\text{post-iDM}}} {\mathcal{E}_{\text{post-FEC}}},
\end{IEEEeqnarray}
which are 200 and 5.5 for CCDM and HiDM, resp., at around $\mathcal{E}_{\text{post-FEC}} = 10^{-6}$. This relation is expected to be valid also at lower error rates, according to the analysis in the next section.

To quantify the amount of burst errors after the invDM, the ratio of BBER $\mathcal{E}_{\text{block}}$ to post-invDM BER $\mathcal{E}_{\text{post-iDM}}$,
\begin{IEEEeqnarray}{rCL}
	\label{eq:rate_BER2}
	r_{\mathcal{E} 2} &=& \left\{
	\begin{matrix}
		\mathcal{E}_{\text{block}} / \mathcal{E}_{\text{post-iDM}} & \text{(CCDM,\,HiDM)} \\
		\mathcal{E}_{\text{block}} / \mathcal{E}_{\text{post-FEC}} & \text{(BICM)} 
	\end{matrix} \, , \right.
\end{IEEEeqnarray}
is useful, which is found to be 500, 11000, and 20000 for CCDM-based PS-256-QAM, HiDM-based PS-256-QAM, and BICM 128-QAM, respectively, as shown in Fig.~\ref{fig:simres}. If $r_{\mathcal{E} 2}$ is smaller, the erroneous frame has a larger number of errors, so the error bursts are more severe. The error burstiness of HiDM-based PS-256-QAM is $11000/500=22$ times lower than with CCDM.
It means that the HiDM can be concatenated with an outer hard-decison FEC if needed because of the significant reduction of the burstiness of the residual errors.

\section{Post-invDM performance estimation at low error rates}
\label{sec:perfmon}
\begin{figure}[tb]
	\begin{center}
		\setlength{\unitlength}{.6mm} %
		\scriptsize
		\vspace{-0.1cm}
		\includegraphics[scale=0.3]{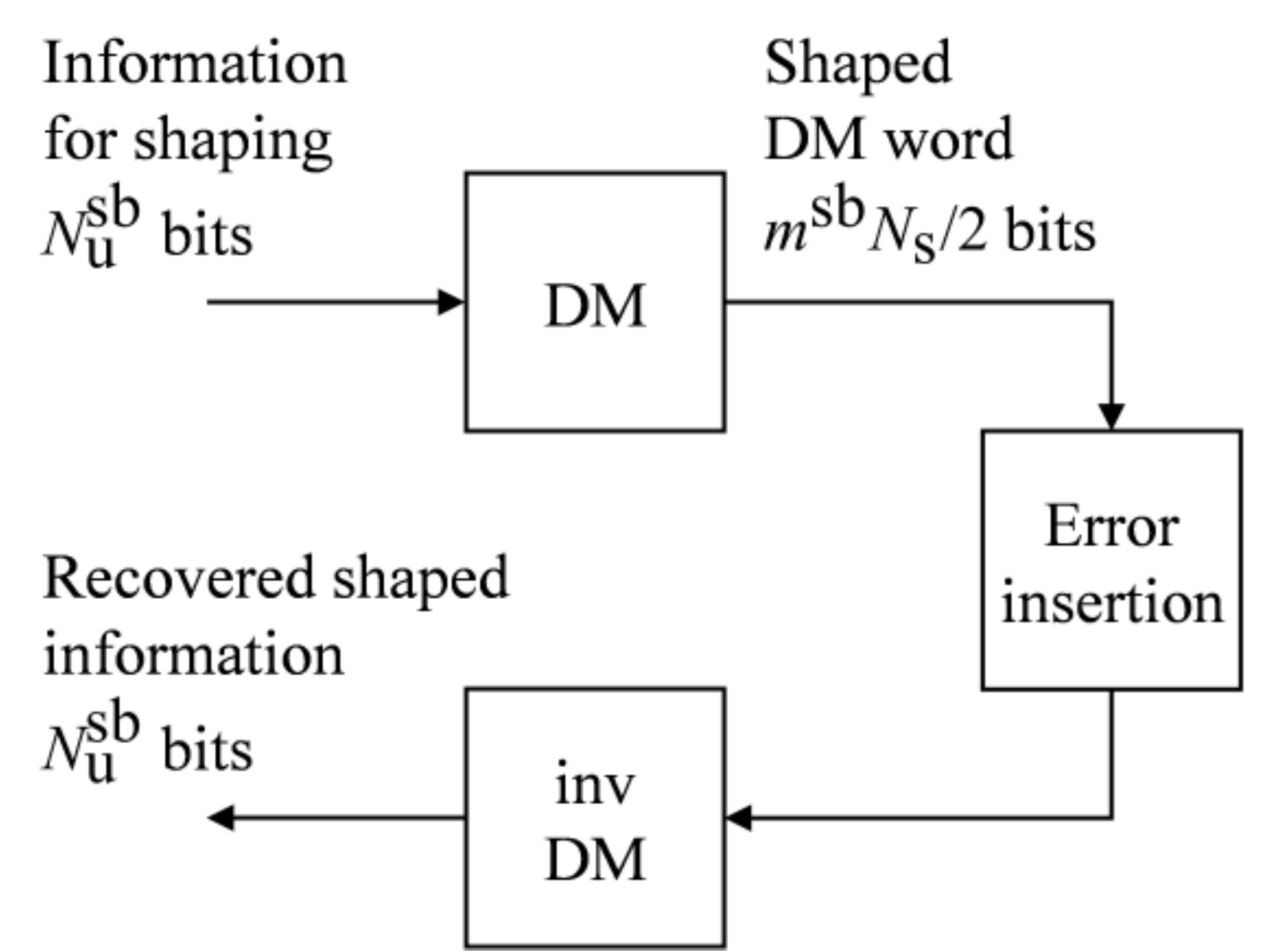} \\
		\vspace{-0.3cm}
		\caption{\uline{Simulation setup for DM to invDM back-to-back error insertion test.}}
		\label{fig:DM_BBsimcnf}
	\end{center}
\end{figure}

We estimated the post-invDM BER from the post-FEC BER by inserting random errors into  each DM word before the invDM operation. Fig.~\ref{fig:DM_BBsimcnf} shows the simulation configuration for the DM to invDM back-to-back error insertion test. As before, the shaped bit levels are $\boldsymbol{i}_{\text{sb}} = \{ 2, 3 \}$ and $m^{\text{sb}}\uline{/2}=2$. The number of information bits $N_{\text{u}}^{\text{sb}} = 1014$ and $507$ for CCDM and HiDM, resp. The information bits were fed to the DM, $m^{\text{sb}} N_{\text{s}}\uline{/2} = 2 N_{\text{s}}$ shaped bits were generated, errors were inserted into the shaped bits, and the invDM recovered the shaped information.

\begin{figure}[tb]
	\begin{center}
		\setlength{\unitlength}{.6mm} %
		\scriptsize
		\vspace{-0.1cm}
		\includegraphics[scale=0.48]{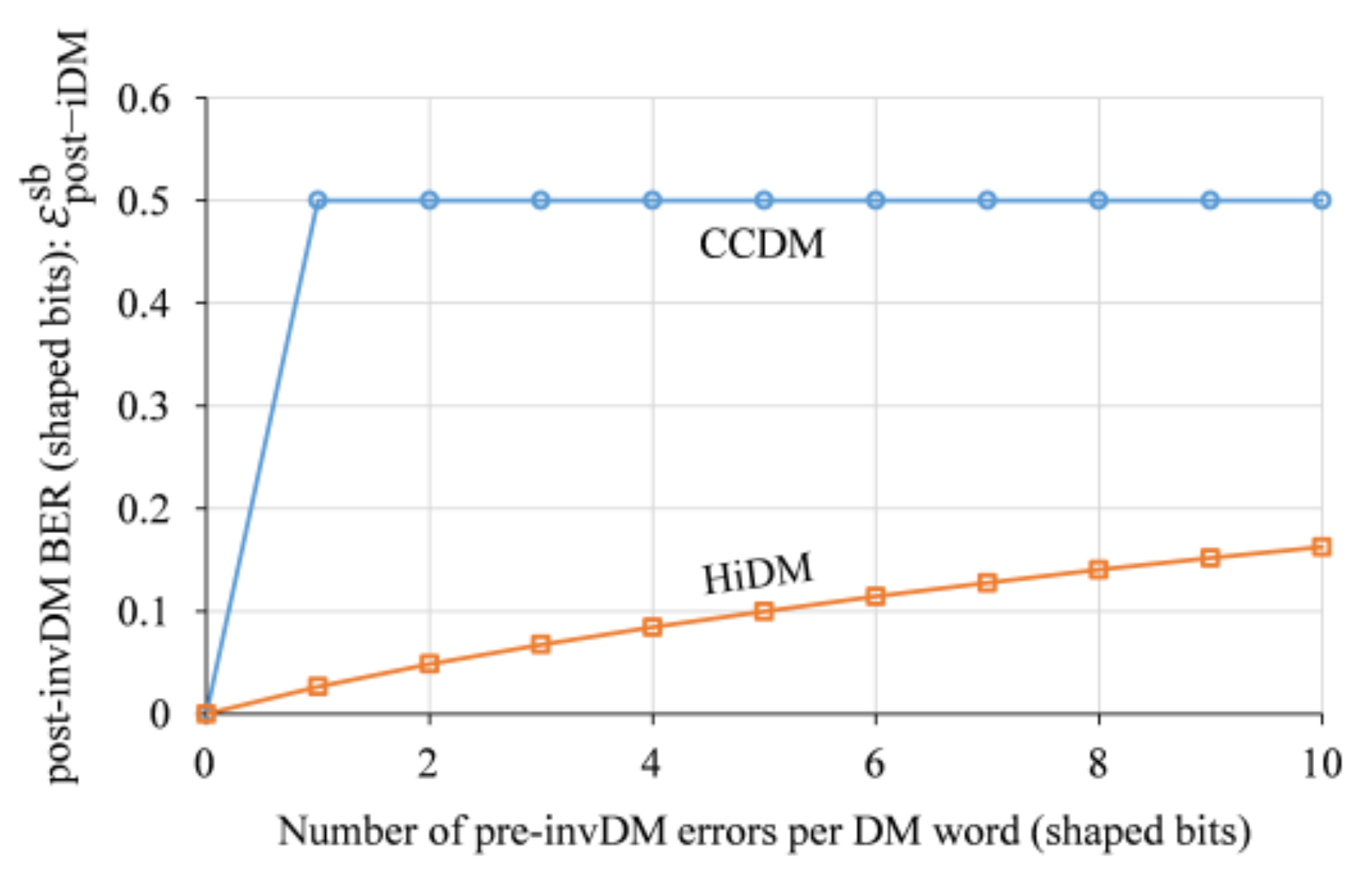} \\
		\vspace{-0.3cm}
		\caption{Simulated post-invDM BER under a single error per DM word at the input of the invDM for PS-256-QAM.}
		\label{fig:DM_BBsim}
	\end{center}
\end{figure}

Fig.~\ref{fig:DM_BBsim} shows the simulated post-invDM BER $\mathcal{E}_{\text{post-iDM}}^{\text{sb}}$ as a function of the number of erroneous bits per DM word, both computed for the shaped bits only.
The expected value of $\mathcal{E}_{\text{post-iDM}}^{\text{sb}}$ for CCDM is $0.5$ for any number of errors, whereas for HiDM it increases from $0.027$ for a single error to $0.16$ for $10$ errors. The expected number of errors after post-invDM $\alpha$ is $507$ for CCDM and $13.4$ for HiDM with a single bit error.


The ratio $r_{\mathcal{E} 1}$ of 200 or 5.5 in the previous section corresponds to the expected error count under the single-error input assumption. Although the FEC decoder can output an error burst, at least the bit-level dependence is weak according to the simulation results in Fig.~\ref{fig:simres}. Under the single-error input condition, 
the post-invDM BERs for the shaped bits and the total bits, $\mathcal{E}_{\text{post-iDM}}^{\text{sb}}$ and $\mathcal{E}_{\text{post-iDM}}^{\text{total}}$, are bounded as
\begin{IEEEeqnarray}{rCL}
	\label{eq:postIDBERsb}
	\mathcal{E}_{\text{post-iDM}}^{\text{sb}} & \le & \text{min} \left\{ \frac{\alpha \cdot m^{\text{sb}} N_{\text{s}} /2} { N_{\text{u}}^{\text{sb}} }  \mathcal{E}_{\text{post-FEC}}, \frac{1}{2} \right\} , \\
	\label{eq:postIDBERtotal}
	\mathcal{E}_{\text{post-iDM}}^{\text{total}} & \le & \gamma_{\text{in}} \mathcal{E}_{\text{post-iDM}}^{\text{sb}} + \left( 1- \gamma_{\text{in}} \right) \mathcal{E}_{\text{post-FEC}} .
\end{IEEEeqnarray}
The fractions of shaped DM input and output bits (i.e., the fractions of $\boldsymbol{A}'$ and $\boldsymbol{D}$ that are shaped in the DM) are, resp.,  
\begin{IEEEeqnarray}{rCL}
	\gamma_{\text{in}} &=& \frac{N_{\text{u}}^{\text{sb}}} {N_{\text{u}}} , \\
	\gamma_{\text{out}} &=& \frac{ m^{\text{sb}} } { R_{\text{c}} m } .
\end{IEEEeqnarray}
The BBER can be bounded as
\begin{IEEEeqnarray}{rCL}
	\label{eq:BBER_bound}
	\mathcal{E}_{\text{block}} & \le & \text{min} \left\{ \frac{N_{\text{block}} \left( \gamma_{\text{out}} \theta + 1 - \gamma_{\text{out}}  \right)} {k \cdot N_{\text{u}} / (R_{\text{c}} m N_{\text{s}} / 2 ) } \mathcal{E}_{\text{FEC-fr}} , 1 \right\} ,
\end{IEEEeqnarray}
where the FEC FER $\mathcal{E}_{\text{FEC-fr}}$ is also bounded as
\begin{IEEEeqnarray}{rCL}
	\label{eq:FECFER_bound}
	\mathcal{E}_{\text{FEC-fr}} & \le & \text{min} \left\{ k \cdot \mathcal{E}_{\text{post-FEC}} , 1\right\} . 
\end{IEEEeqnarray}
The parameters $\alpha$ and $\theta$, which denote the number of output error bits of the invDM with a single-error input and the number of OTUC1 block errors with a single FEC codeword error, resp., are shown in Tab.~\ref{tab:err_param}, along with $\gamma_{\text{out}}$ and $\gamma_{\text{in}}$.
In the simulations, the number of information bits for the FEC is $k=54000$. If the expected number of errors in an erroneous FEC codeword is larger than 1 due to the bursty nature of FEC decoders, $\mathcal{E}_{\text{FEC-fr}}$ becomes closer to $\mathcal{E}_{\text{post-FEC}}$.
The quantity $N_{\text{u}} /$ \uline{$(R_{\text{c}} m N_{\text{s}}/2)$} \sout{$(m N_{\text{s}} R_{\text{c}} )$} in \eqref{eq:BBER_bound} is 0.875 for CCDM/HiDM-based PS-256-QAM and $1$ for BICM-based 128-QAM.

\begin{table}[t]
	\caption{Single-error parameters}
	\label{tab:err_param}
	\begin{center}
		\begin{tabular}{cccc}
			\hline \hline
			& CCDM & HiDM & BICM \\ \hline
			$\alpha$ & 507 & 13.4 & $-$ \\
			$\theta$ & 32 & $\le$2 & 1 \\
			$\gamma_{\text{in}}$ & 0.543 & 0.543 & 0 \\
			$\gamma_{\text{out}}$ & 0.6 & 0.6 & 0 \\
			\hline \hline
		\end{tabular}			
	\end{center}
\end{table}



The ratio $r_{\mathcal{E} 1}$ can be calculated based on (\ref{eq:postIDBERsb}) and (\ref{eq:postIDBERtotal}), which gives $r_{\mathcal{E} 1}=348$ and $9.7$ for CCDM and HiDM, resp. These theoretic results are of the same order as the simulation results in the previous section. 
This single-error scenario gives the largest $r_{\mathcal{E} 1}$ (the post-invDM BER would be worse than the case with burst errors). 

\begin{figure}[tb]
	\begin{center}
		\setlength{\unitlength}{.6mm} %
		\scriptsize
		\vspace{-0.1cm}
		\includegraphics[scale=0.48]{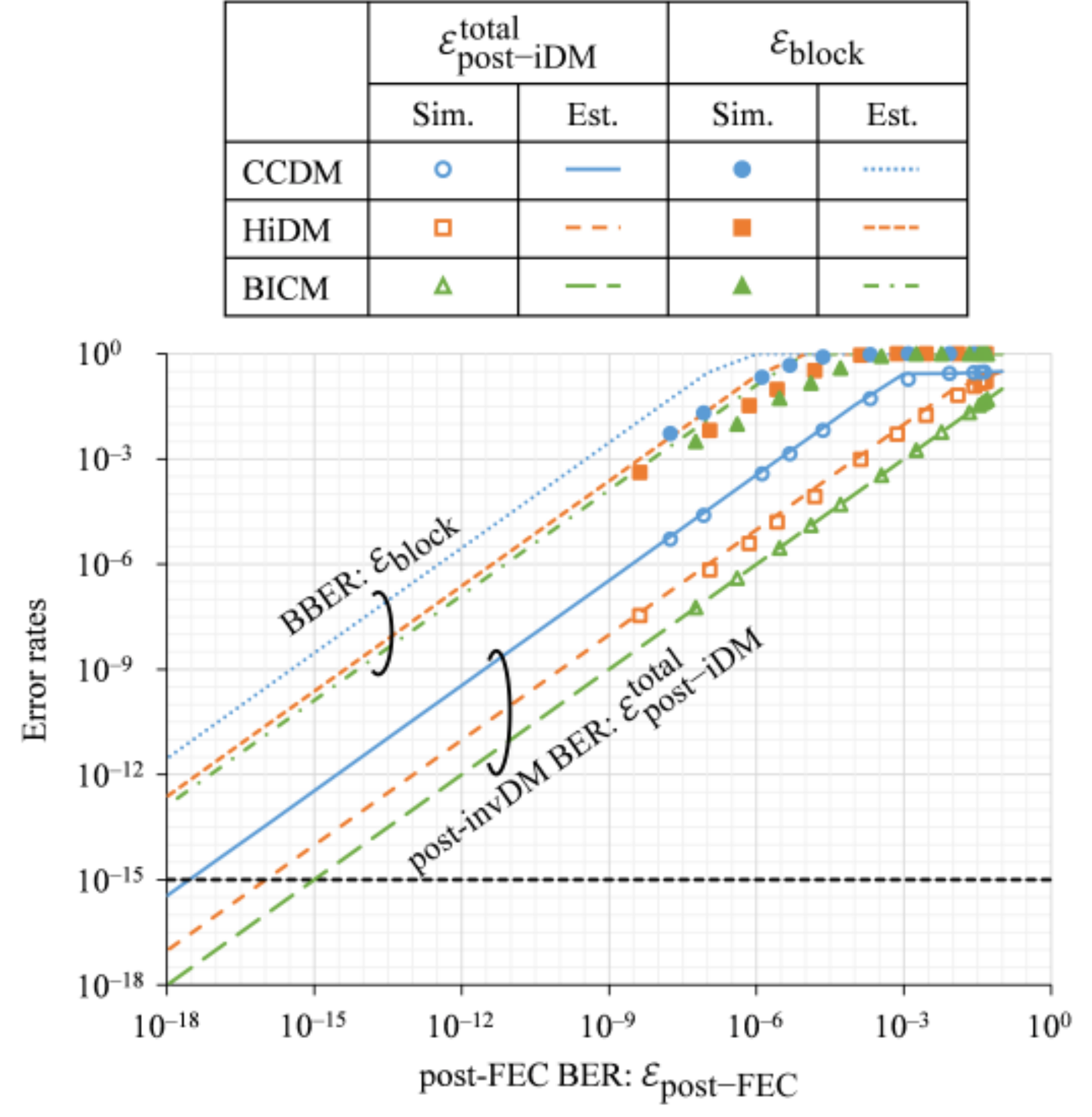} \\
		\vspace{-0.3cm}
		\caption{Post-invDM BER from simulation or estimation under the assumption of a single-error input to the invDM per DM word.}
		\label{fig:est_postIDBER_low}
	\end{center}
\end{figure}

In Fig.~\ref{fig:est_postIDBER_low}, we estimated the post-invDM BER and BBER as a function of the post-FEC BER at low error rates using the bounds (\ref{eq:postIDBERtotal}) and (\ref{eq:BBER_bound}). 
The post-invDM BER $\mathcal{E}_{\text{post-iDM}}^{\text{total}}$ estimates agree well with the simulation results.
To achieve a post-invDM BER of $10^{-15}$, a post-FEC BER of $3\cdot10^{-18}$ (for CCDM) or $10^{-16}$ (for HiDM) is required. We can thus conclude that the FEC design and evaluation becomes more difficult with reverse concatenation PS than with BICM, due to the requirement of such low post-FEC BERs. If the code would have an error floor just below a post-FEC BER of $10^{-15}$, the required SNR for a post-invDM BER of $10^{-15}$ would be significantly larger. Also, simulations to evaluate a post-FEC BER of $3\cdot10^{-18}$ to ensure that there is no error floor are very challenging (to transmit $10^{18}$ bits at 100 Gbit/s takes $10^7$ s, or 3.8 months). HiDM requires an order of magnitude higher error rate than CCDM, so the gap between BICM and reverse concatenation PS is greatly reduced.
As for the BBER $\mathcal{E}_{\text{block}}$, estimations and simulations approach each other in the low post-FEC BER regime, where the single-error assumption works well. The bounded BBER as a function of the post-FEC BER for HiDM is almost the same as for BICM-based 128-QAM. The BBER bound for HiDM is 16 times lower than for CCDM in the case of the same post-FEC BER, because the HiDM places the DM word on the local area due to the high throughput feature compared with the CCDM, as shown in Fig.~\ref{fig:frame}. 
\uline{When the DM word length $N_{\text{s}}$ is increased, the post-FEC error rates are reduced,  but the bounds on $\mathcal{E}_{\text{post-iDM}}^{\text{sb}}$ and $\mathcal{E}_{\text{block}}$ in \eqref{eq:postIDBERsb} and \eqref{eq:BBER_bound} increase proportionally to $N_{\text{s}}$.}


\section{Summary and outlook}
\label{sec:cncl}
We proposed HiDM for low-complexity implementation of reverse-concatenation PS, and evaluated its end-to-end performance for high-throughput fiber-optic communications.
As the post-invDM BER and BBER performance depends on the FEC decoding, invDM processing, and cross-layout of the FEC codewords and the DM words, we demonstrated frame structure examples 
considering the combined FEC and PS. 
By a back-to-back error insertion test and a single-error assumption, very low post-invDM error rates can be estimated from the post-FEC BER, which allows relaxed requirements for the post-FEC BER if HiDM is used instead of CCDM. 
The combination of known post-FEC BER prediction techniques using NGMI or ASI with the proposed post-invDM error rate estimation from post-FEC BER would be an interesting topic for further studies.
Other future research directions include a careful design of DM/invDM algorithms considering the implementation combined with FEC to reach the desired BER and BBER performance targets.

\section*{Acknowledgments}
We thank Alex Alvarado of Eindhoven University of Technology\uline{, David S. Millar of Mitsubishi Electric Research Laboratories,} and Koji Igarashi of Osaka University for fruitful discussions about performance metrics\uline{, distribution matching,} and reverse concatenation.



\end{document}